\documentclass[aps,prl,nofootinbib,twocolumn, groupedaddress,letterpaper,amsmath,amssymb]{revtex4}
\usepackage{xcolor, mdframed}
\usepackage{graphicx}
\usepackage{natbib}
\usepackage{color}
\usepackage{subcaption}
\usepackage{float}
\usepackage{hyperref}
\usepackage{wrapfig}
\usepackage{amssymb}
\usepackage{mathtools}
\usepackage[rightcaption]{sidecap}
\usepackage{stfloats}
\usepackage{float}

\DeclarePairedDelimiter\abs{\lvert}{\rvert}%
\DeclarePairedDelimiter\norm{\lVert}{\rVert}%

\makeatletter
\let\oldabs\abs
\def\abs{\@ifstar{\oldabs}{\oldabs*}}
\let\oldnorm\norm
\def\norm{\@ifstar{\oldnorm}{\oldnorm*}}
\makeatother

\begin{document}

\title{Finding Unexpected Non-Helical Tracks}
\author{Levi Condren}
\author{Daniel Whiteson}
\affiliation{Department of Physics \& Astronomy, University of California, Irvine, CA}

\begin{abstract}
Many theories of physics beyond the Standard Model predict particles with non-helical trajectories in a uniform magnetic field, but standard tracking algorithms assume helical paths and so are incapable of discovering non-helical tracks.  While alternative algorithms have been developed for specific trajectories,  unforeseen physics could lead to unanticipated behavior, and such unexpected tracks are largely invisible to current algorithms, despite being potentially striking to the naked eye.  A model-agnostic tracking algorithm is presented, capable of reconstructing a broad class of smooth non-helical tracks without requiring explicit specification of particle trajectories, instead defining the target trajectories implicitly in the training sample. The network exhibits strong performance, even outside of the trajectories defined by the training sample. This proof-of-principle study takes the first step towards searches for unexpected tracks which  may await discovery in current data.
\end{abstract}

\maketitle

\section{Introduction}
\label{sec:introduction}
Particle colliders such as the Large Hadron Collider (LHC) have tremendous potential to discover new physics \cite{ATLAS:2012yve,CMS:2012qbp,annurev:/content/journals/10.1146/annurev-nucl-101920-014923,annurev:/content/journals/10.1146/annurev-nucl-102419-052854,LHCb:2015yax,LHCb:2014vgu,ATLAS:2010isq,CMS:2010ifv}, but require computationally expensive analysis to take full advantage of their data. One of the most challenging tasks is to identify a set of detector responses due to an individual charged particle, a {\it track}, among the enormous number left by all particles produced in a collision.  Considering every possible subset 
is  computationally intractable task due to the vast combinatorics in the most general case~\cite{Strandlie:2010zz}. Traditionally, simplifying assumptions are employed to make the task feasible~\cite{ATLAS:2023iat}. Primary among these is restricting the search to tracks which follow the helical trajectory expected of electrically-charged Standard Model particles in a uniform magnetic field. This greatly simplifies the track identification task, but  leaves experiments incapable of finding non-helical tracks, shrinking the potential for discovery.

Several theories of physics beyond the Standard Model (BSM) have interactions which produce non-helical trajectories, such as magnetic monopoles~\cite{ATLAS:2019wkg, CDF:2005cvf} or quirks~\cite{Kang:2008ea, Evans:2018jmd, Knapen_2017}.  Recent work on quirks ~\cite{Sha:2024hzq} has demonstrated the capability to efficiently find non-helical tracks when the predicted trajectories are specified in advance.
  However, the space of possible particle trajectories is much larger than what is described by the current set of BSM theories. {\it Model-agnostic} tracking, capable of finding tracks without a pre-existing theory or pre-specified trajectory, would broaden experimental capacity for unexpected discoveries, especially in low-background, high-fidelity environments such as tracking detectors~\cite{PhysRev.43.491}.  Tracks with striking, surprising trajectories may exist in current experimental datasets, undiscovered only because they are invisible to current algorithms.

 Most traditional track-finding algorithms~\cite{ATLAS:2017kyn}  perform alternating track parameter fitting and iterative hit finding, which requires knowledge of the explicit parametric form of the track, and so are fundamentally unsuited to model-agnostic tracking. However, recent developments  in machine learning such as graph neural networks (GNNs)~\cite{Bronstein:2016thv} allow for track-finding algorithms where the form of the track is never specified explicitly, only implicitly via the sample used to train the algorithm~\cite{ExaTrkX:2020nyf,Ju_2021}. While not the original goal of ML tracking, this new approach  separates the finding and fitting steps~\cite{Caillou:2024smf, Caillou:2022hly}, and therefore allows for finding non-helical trajectories when trained on such examples~\cite{Sha:2024hzq}. Crucially, it may also open the door to model-agnostic tracking, if the broad space of possible trajectories can be described implicitly via a training sample, and the  tracker learns to generalize beyond the specific training examples.


In this paper, a GNN-based physics-model-agnostic track finder is trained on a sample of smooth tracks generated without a specific trajectory model via selection of Fourier amplitudes in frequency space.
Smoothness is guaranteed by imposing a Schwartz condition on the relationship among Fourier amplitudes. The performance of the track finder to reconstruct non-helical smooth tracks which are similar to those in its training sample is measured, as well as its ability to generalize to tracks dissimilar to those in its training sample.

In Section~\ref{sec:modeling}, the simulation of tracks is described, including the detector geometry and model-agnostic track generation.  In Section~\ref{sec:tracking}, the tracking pipeline is described. Section~\ref{sec:tracking} details the reconstruction performance on various sets of tracks, including background rejection and tests of generalization. Section~\ref{sec:discussion} discusses the results and next steps.


\section{Modeling}

Simulated tracks are used to train the track-finding pipeline, and to assess  reconstruction efficiency, generalization performance, and rate of background tracks. Below,  the simulation of the detector response, generation of helical standard model (SM) background tracks and generation of non-helical tracks are described. Training and testing samples contain combinations of many helical SM tracks and at most one non-helical track.

\label{sec:modeling}

\subsection{Detector Geometry}
A custom detector geometry is used, with  25 coaxial cylindrical layers with a length of 320 cm and even radial spacing ranging between 3.1 cm and 53 cm. Tracks are generated by producing trajectories which originate at the center of the detector. 

Hits are recorded at every intersection of each trajectory with a detector layer until the particle's path leaves the detector, because the pipeline is trained to find sets of hits that correspond to continuous trajectories which remain within the volume.   Many trajectories do return to the detector and leave additional hits, effectively producing several disconnected sets of hits from a single trajectory. While piecing together several such sets may give additional discovery power, it also raises a new and distinct set of reconstruction challenges, and so is left to future work.

Particle energy loss and hit location uncertainty are neglected, but similar work did not show strong sensitivity~\cite{Sha:2024hzq}.

\subsection{SM Background} \label{subsec:sm}
Madgraph5 \cite{Alwall:2014hca} is used to generate $pp \rightarrow t \bar{t}$ events, which are showered and hadronized using Pythia8 \cite{Bierlich:2022pfr}. Initial velocity and momenta are used to propagate Standard Model particles in helical trajectories under the assumption that there is no energy loss or secondary scattering. 

 All events use a pileup of $\mu = 1$. The impact of  pile-up on performance is not yet known, but in other studies the same GNN pipeline performed well in high pile-up conditions similar to those expected at the high-luminosity LHC~\cite{ExaTrkX:2021abe}. Such performance may extend to the case of generalized smooth tracks, or it may be more susceptible to confusion in such hit dense environments. A rigorous study is out of scope for this initial proof-of-principle.

\begin{figure}
    \centering
    \includegraphics[width=0.6\linewidth]{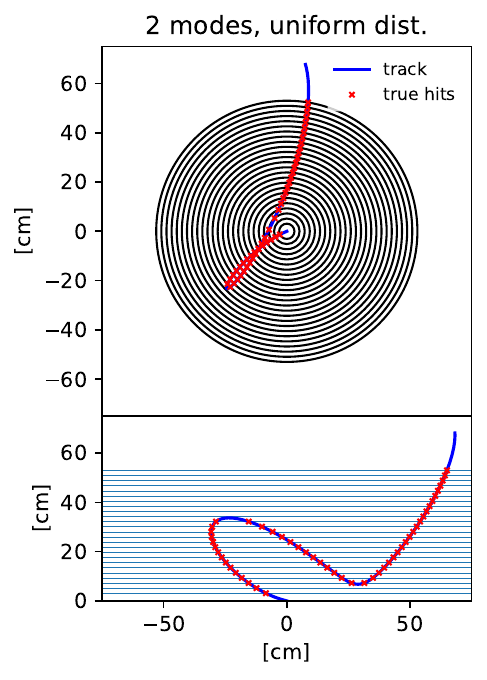}
    \includegraphics[width=0.6\linewidth]{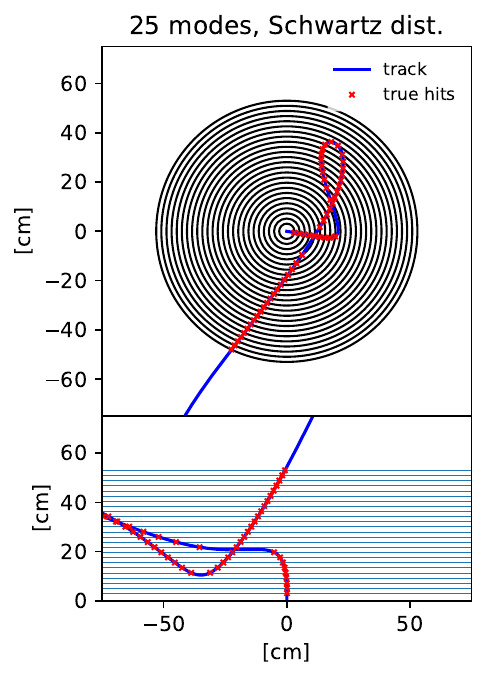}\\
    \caption{Trajectories (blue) and hits (red) for tracks with 1 (top left), 2 (top right),  4 (bottom left), or 25 (bottom right) Fourier modes, where the 25-mode track follows a Schwartz function to ensure smoothness. Shown are the $x-y$ (top) and $r-z$ projections.}
    \label{fig:trackviz}
\end{figure}

\subsection{Non-Helical Track Generation} \label{subsec:sig}

The model-agnostic search aims to be sensitive to a very broad range of possible new particles with unexpected trajectories through the detector. However, it cannot accept every possible trajectory, or all sets of hits would be  candidates. Instead, particle tracks are required to follow  {\it smooth} paths, those which can be parameterized by a curve, $\vec{x}(t)$ where all orders of spatial derivatives with respect to time, $\frac{d^nx}{dt^n}$, exist and are finite, which accepts all particles with physical paths, where the net force and its derivatives, $m  \frac{d^n \ddot{x}}{dt^n}$,  are physical\footnote{While this may be untrue for idealized or simplified systems which exhibit discontinuities in external forces, fully physical classical systems disregarding quantum effects such as decays, are subject to classical field interactions, which are infinitely differentiable with respect to position and time \cite{Landau:1975pou}.}. It rejects candidates whose paths are discontinuous or kinked. This method of track generation narrows our analysis by disregarding non-prompt decays, and particles that scatter off of detector material. While many particles exhibit non-smooth behavior in detector environments, a smoothness assumption provides a theoretically sound foundation for demonstrating this proof-of-concept analysis.  
\\

Without  assumption about the underlying physical theory, three-component spatial curves can be fully described in a Fourier basis by three sequences of Fourier amplitudes, $(a_n,b_n,c_n)$ defined in a Cartesian basis.  Such paths, however, are not necessarily smooth unless each sequence approaches zero as $n \rightarrow \infty$ at least as fast as a Schwartz function $f(n)$\cite{Stein:2003,Hormander:1990,Reed:1980}. Explicitly,  smoothness is guaranteed if $a_n, b_n, c_n \le f(n)$ for all $n$. 

Figure \ref{fig:trackviz} shows example trajectories generated with a small number of frequencies, as well as one following Schwartz function 1 defined in Tab.~\ref{tab:sf}.

\begin{figure}[h]  
    \centering
    \includegraphics[width=0.48\textwidth]{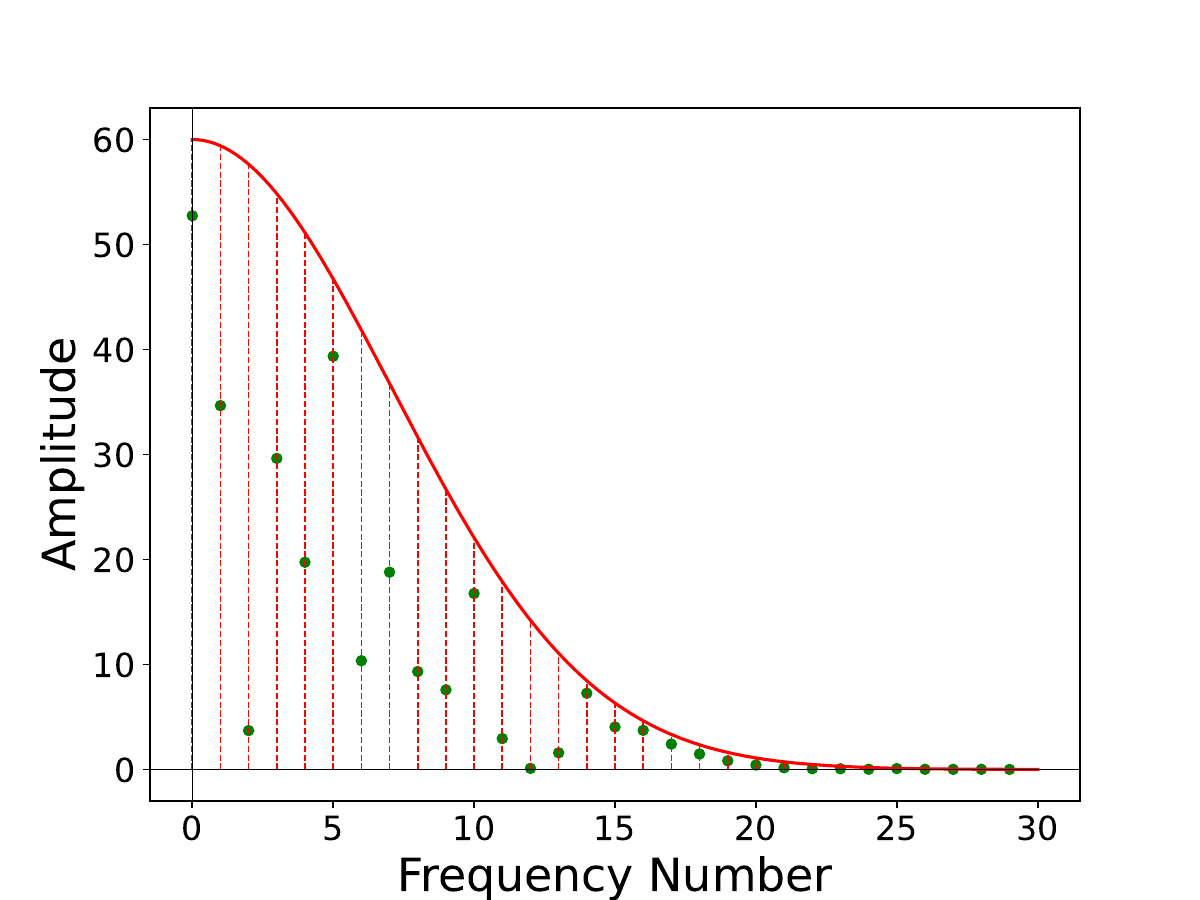}
    \caption{ An example of a Schwartz function (red), which enforces a falling upper limit on amplitude as frequency grows, guaranteeing smoothness if amplitudes are selected (green dots) below the upper limit.}
    \label{fig:sf}    
\end{figure}

Figure~\ref{fig:sf} provides a visual example of generating a smooth amplitude by limiting amplitudes at higher frequencies. Note that a three dimensional path requires three such curves, one for each coordinate. 

While all trajectories generated using a Schwartz function limit are smooth, and for any smooth trajectory there is a Schwartz function that would allow its generation (proofs in App. \ref{appendix:shwartz}), in principle describing the full space of smooth tracks may require an infinite number of frequency modes and many Schwartz functions.  In practice, sampling from infinitely many frequencies is not computationally feasible, but Schwartz functions' rapid decay make high-frequency modes nearly vanish. Sampling from a large number of frequencies, 25-30, therefore explores a significant fraction of the space. Particle displacement resulting from frequencies beyond this would contribute negligible oscillatory motion, much less than the radial spacing of detector layers.


Similarly, a set of Schwartz functions are selected to explore the space, which does not span the full, infinite space, but  describes tracks significantly beyond the set of helical tracks. The chosen Schwartz functions include helices and explicit parametrization of helical tracks in a Fourier basis is shown in App. \ref{appendix:shwartz}.




For demonstration, non-helical tracks with a small number of frequencies are also prepared, by sampling from finitely many frequencies in Fourier space, where each mode has a maximum amplitude of 60cm. An $N$-frequency track is defined as the collection of points where a function $\vec{g}_N({t})=\sum_{i=1}^3 \sum_{n=-N}^N c_{ni}\exp(\frac{2 \pi i n t}{\lambda})\hat{x_{i}}$ intersects the detector layers, where $\hat{x_{i}}$ is a cartesian basis, and the $c_{ni}$ are all randomly generated with a uniform distribution in the interval $[0, 60]$ cm, and $\lambda$ is set equal to the radius of the detector which is a choice further discussed in App. \ref{sig-gen}.

\begin{figure}[h]
    \centering
    \includegraphics[scale=0.5]{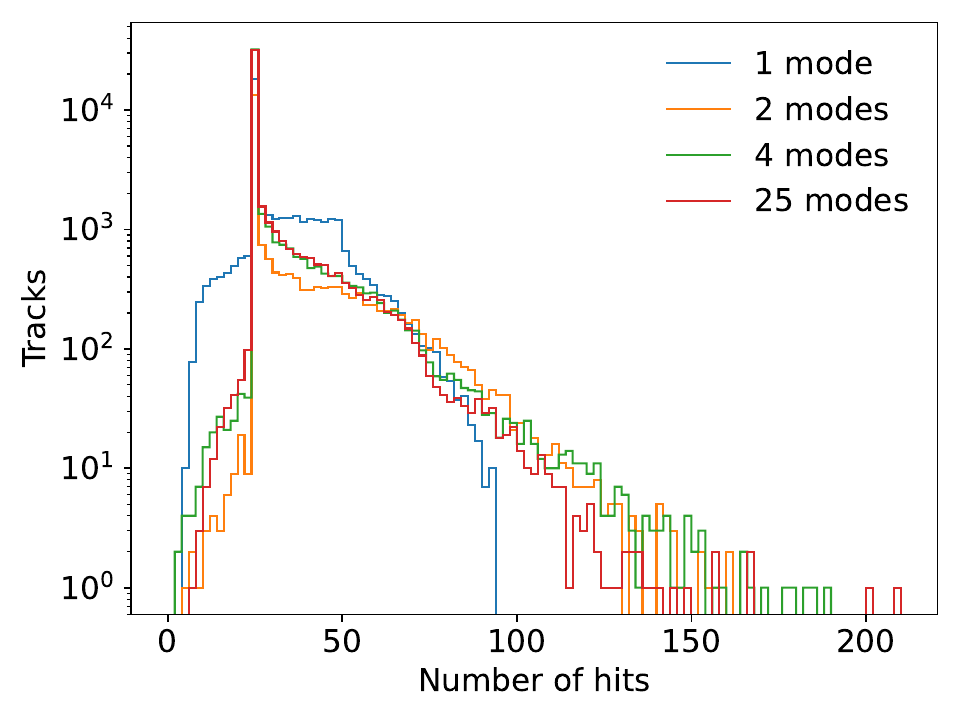}
    \caption{Shown are the distributions of number of hits per generated track depending on the number of frequency modes used in the trajectory generation.} 
    \label{fig:hits}
\end{figure}

The distribution of the hit multiplicity for tracks with varying number of frequencies is shown in Figure~\ref{fig:hits}; while a significant number have as many hits as there are detector layers, a large fraction have many more hits.

\section{Tracking}
\label{sec:tracking}

Track reconstruction involves two distinct steps, {\it finding}, where a set of hits are identified as a track candidate, and  {\it fitting}, which finds best values of parameters in an assumed form to minimize track-to-hit distance.  Reconstructed tracks are crucial elements of data analysis in many areas, including triggers, vertex finding and jet flavor tagging.

Many tracking algorithms, such as a Kalman Filter, assume a parametric track path and ionization and interaction properties typical of known particles. An initial set of hits allows for initial estimates of the track parameters, which in turn allows for propagation of the track to subsequent detector surfaces and a calculation of a window within which additional hits should be considered. This hit-by-hit approach naturally weaves together track finding and fitting. Machine learning can be used in a variety of ways to perform track finding\cite{duarte2022graph,dezoort2023graph,thais2022graph}. For example, ML can be used to propagate initial tracks in a Kalman Filter~\cite{heinrich2024combined}, or more broadly to select the next hit~\cite{vaage2022reinforcement,kortus2023towards}. 

The machine-learning-based approach used below does not require an explicit model of the trajectory. Instead, the relationships between hits on the same track are implicitly defined through the example tracks in the training sample. A track is represented as a directed graph, with hits as nodes. Track-finding pipelines, such as {\sc Exa.TrkX}, which use  graph neural networks achieve high track-finding efficiencies in a broad variety of particle experiments, even under high pile-up conditions~\cite{caillou2022atlas,verma2020particle,hewes2021graph,drielsma2021scalable,akram2022track,jia2024besiii,biscarat2021towards,zdybal2024machine,andrews2021accelerating,dezoort2021charged,Correia_2024,Elabd_2022}. The {\sc Exa.TrkX} pipeline first performs graph construction, where hits are linked into a graph, followed by edge classification, where graph edges are evaluated as their likelihood to connect two hits which belong to the same track, followed by graph segmentation, where edges are pruned and collected hit connections form track candidates. The {\sc Exa.TrkX} pipeline employed here relaxes the requirement that hits are ordered in time according to their distance from the vertex, following Ref.~\cite{Sha:2024hzq}. Graph edges are evaluated by a metric learning model, which learns a mapping from the physical space of hits to an abstract latent space, such that hits on the same track are near each other and hits from other tracks are more distant.

The {\sc Exa.TrkX} pipeline employed requires information about the hit ordering. Traditionally, the assumption is made that hit ordering follows the layers from inside out.  As tracks may change direction, in this application the true hit ordering is used following Ref.~\cite{Sha:2024hzq}. For SM tracks, this has little impact as backscattering toward the production vertex occurs infrequently, and the fake rate remains low.

The metric learning model is trained for 5 epochs using on the order of $10^4$ training events with a learning rate of 0.001. The edge classification model is trained for another 5 epochs, using early stopping and variable learning rate to avoid overfitting the data. 

The primary performance metric is the tracking efficiency, the fraction of {\it reconstructable} particles (those with at least 22 hits) that are {\it reconstructed}, double-majority matched to a {\it matchable} track candidate (those with at least 15 hits). A track is double-majority matched to a track candidate when the majority of hits in the reconstructed track belong to the same truth track, and the majority of hits in the truth track belong to the reconstructed track. Tracks are labeled as {\it fake} if they do not double majority match any SM or non-helical track.   The pipeline may also be capable of reconstructing shorter tracks; such studies are reserved for future work. 

\begin{table}[h!]
    \centering
    \caption{Shown is track finding efficiency for non-helical tracks in various testing and training configurations. SM refers to helical tracks, $n$ Freq refers to tracks generated with $n$ Fourier modes, and SS refers to tracks generated with 25 Fourier modes following a specific Schwartz function.}
      \label{tab:sm-sm+sig}
    \begin{tabular}{ll|rr}
    \hline\hline
    Train & Test & Efficiency & Fake Rate \\ 
    \hline
    SM & SM + Signal \\
    \hline
    SM & SM + 1 Freq.  & 0.028 $\pm$ 0.008 & 0.383 \\
    SM & SM + 2 Freq.  & 0.019 $\pm$ 0.006 & 0.019 \\
    SM & SM + 3 Freq.  & 0.017 $\pm$ 0.004 & 0.391 \\
    SM & SM + 4 Freq.  & 0.029 $\pm$ 0.002 & 0.200 \\
    SM & SM + SS 1  & 0.034 $\pm$ 0.009 & 0.423 \\
    \hline
    Signal & Signal \\
    \hline
    1 Freq.  & 1 Freq.  & 0.964 $\pm$ 0.006 & 0.001 \\ 
    2 Freq.  & 2 Freq.  & 0.962 $\pm$ 0.004 & 0.000 \\
     3 Freq.  & 3 Freq.  & 0.983 $\pm$ 0.002 & 0.001 \\
     4 Freq.  & 4 Freq.  & 0.993 $\pm$ 0.002 & 0.001 \\
     SS 1  & SS 1  & 0.993 $\pm$ 0.001 & 0.000 \\
    \hline 
    SM + Signal & SM + Signal \\
    \hline 
     SM + 1 Freq.  & SM + 1 Freq.  & 0.475 $\pm$ 0.020 & 0.002\\ 
     SM + 2 Freq.  & SM + 2 Freq.  & 0.561 $\pm$ 0.013 & 0.002\\
    SM + 3 Freq.  & SM + 3 Freq.  & 0.571 $\pm$ 0.011 & 0.001\\
    SM + 4 Freq.  & SM + 4 Freq.  & 0.626 $\pm$ 0.009 & 0.001\\
     SM + SS 1  & SM + SS 1  & 0.589 $\pm$ 0.009 & 0.001 \\
    \hline\hline
    \end{tabular}
\end{table}

\begin{table}[h!]
    \centering
      \caption{Shown is the track finding efficiency for non-helical tracks when training and testing are performed on disjoint spaces as described in Section~\ref{subsec:gen}. Training and testing both have non-helical tracks and SM background}
      \label{tab:generalization-test}
    \begin{tabular}{ll|rr}
    \hline\hline
    Train & Test & Efficiency & Fake Rate \\ 
    \hline

    SM + SS 1 & SM + SS 2-1 & 0.775 $\pm$ 0.007 & 0.000\\ 
    SM + SS 3 & SM + SS 4-3 & 0.668 $\pm$ 0.008 & 0.001\\
    SM + SS 1 & SM + SS 8-1 & 0.667 $\pm$ 0.009 & 0.001\\
    SM + SS 3 & SM + SS 2-3 & 0.592 $\pm$ 0.008 & 0.008\\
    SM + SS 1 & SM + SS 9-1 & 0.590 $\pm$ 0.009 & 0.001\\
    SM + SS 11 & SM + SS 8-11 & 0.579 $\pm$ 0.009 & 0.001\\
    SM + SS 4 & SM + SS 2-4 & 0.546 $\pm$ 0.010 & 0.011\\
    SM + SS 7 & SM + SS 4-7 & 0.497 $\pm$ 0.014 & 0.001\\
    SM + SS 1 & SM + SS 4-1 & 0.414 $\pm$ 0.010 & 0.002\\
    SM + SS 5 & SM + SS 6-5 & 0.008 $\pm$ 0.002 & 0.002\\
    \hline\hline
    \end{tabular}
\end{table}

    
\subsection{Baseline tests}
\label{subsec:base}

The expected performance of the pipeline is verified on SM tracks, training on 40k SM $pp \rightarrow t \bar{t}$ events, and testing on 4k SM events. The reconstruction efficiency is 99.9\% and the fake rate is consistent with zero.  

To estimate of the difficulty of finding non-helical tracks with traditional helical approaches, the ability of the SM-trained pipeline to reconstruct our non-helical tracks is measured in events with one non-helical track added to an SM event. Efficiencies on the non-helical tracks are given in Table~\ref{tab:sm-sm+sig}, and are quite poor, indicating as expected that helical track finders are not well suited to finding non-helical tracks.



The ML-based tracking pipeline has no inherent preference for helical tracks, and should allow for non-helical track finding simply by preparing the appropriate training sample.  A non-helical track finder is trained with a sample of pure non-helical tracks, one per event. Training and testing on such events achieves a very high efficiency (see Tab~\ref{tab:sm-sm+sig}), demonstrating that the pipeline has the capacity in principle to learn the patterns of hits on a wide variety of non-helical tracks.

In practice, non-helical tracks are expected to be found among many SM tracks. In the next stage, SM tracks are added to the training sample and given the same labels\footnote{In previous work which attempted to learn one specific non-helical track pattern~\cite{Sha:2024hzq}, SM tracks were labeled as fake to suppress their reconstruction. In contrast, here the aim is to learn all possible smooth tracks.} Efficiency is lower in the presence of many background tracks, as expected, but remains strong; see Tab.~\ref{tab:sm-sm+sig}.  Even in the presence of SM backgrounds, the tracking pipeline can efficiently find non-helical tracks.

\subsection{Generalization tests}
\label{subsec:gen}

The training sample is necessarily finite, and is constructed using a finite number of Fourier modes from one of many Schwartz functions. An important question is whether the tracker has learned the general properties of smooth tracks, allowing it to generalize outside of its training sample.  Previous work~\cite{Sha:2024hzq} showed hints that the tracking pipeline may be capable of generalization, learning to reconstruct tracks which fall outside of the space described by the training set.

\begin{figure}[h]
    \centering
    \includegraphics[scale = 0.45]{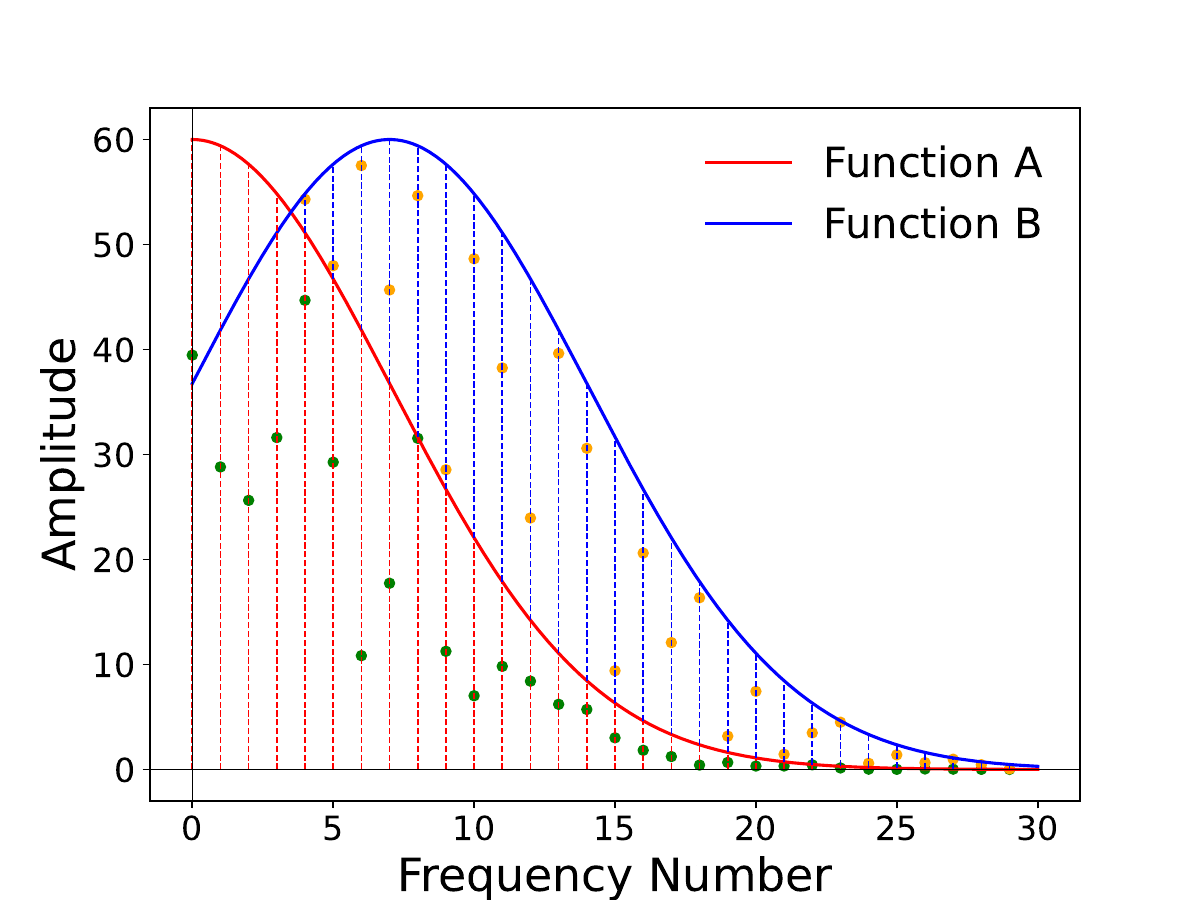}
    \includegraphics[scale = 0.45]{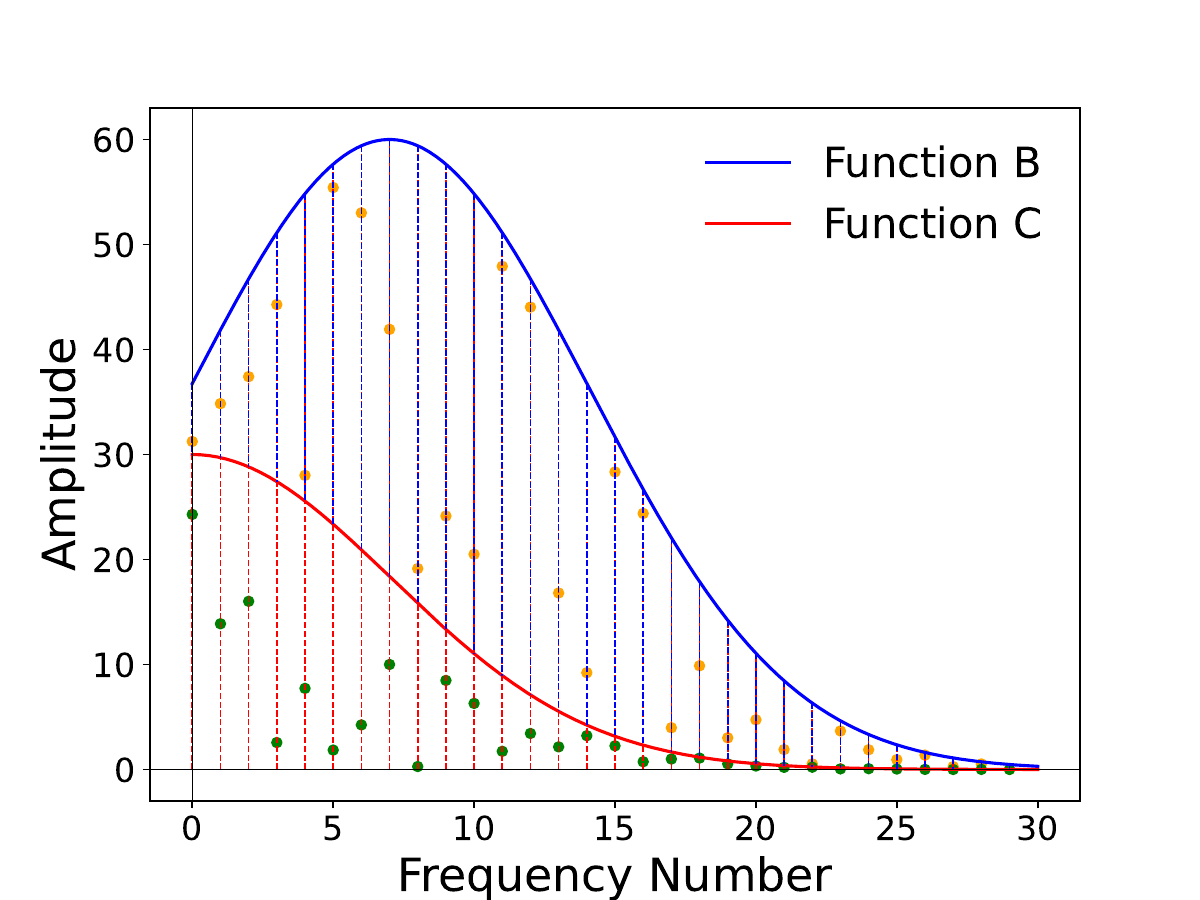}
    \caption{Visualization of tracks generated in disjoint spaces defined by three example Schwartz functions A,B,and C. In green are Fourier coefficients generated under the red curve. In yellow are the coefficients generated between the blue and red curves. The two resulting tracks cannot have the same Fourier representation, and must be different.}
    \label{fig:manysf}
\end{figure}

\begin{figure}[h]
    \centering
    \includegraphics[width=0.75\linewidth]{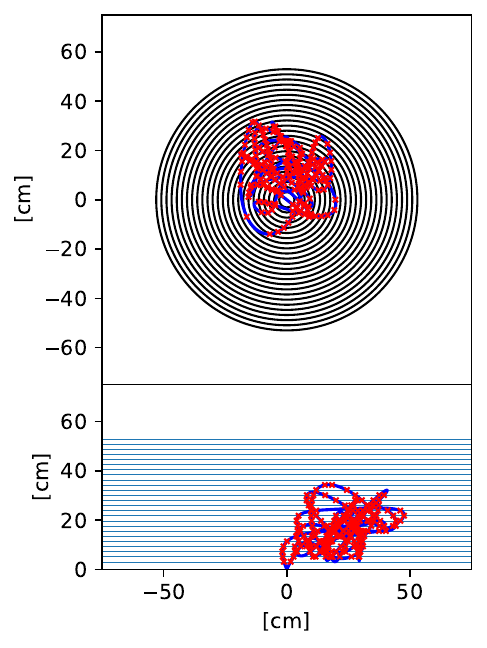}
    \includegraphics[width=0.75\linewidth]{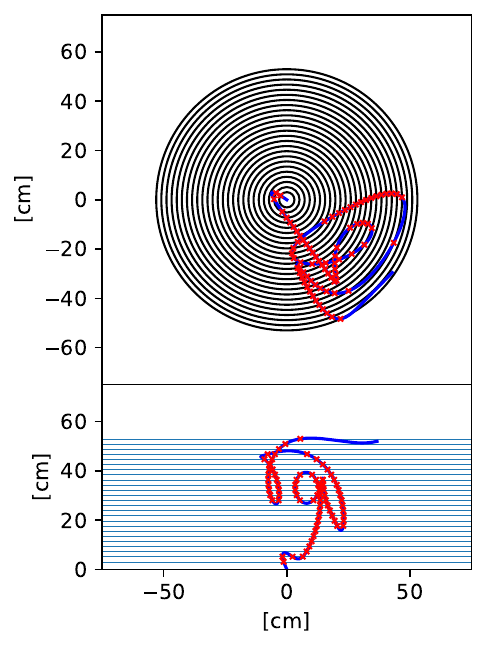}
    \caption{Examples of  tracks generated in the 6-5 space defined in the text.}
    \label{fig:6-5}
\end{figure}

To measure the capacity for generalization, a different Schwartz function is chosen for the training and testing sets. The testing set includes only the disjoint subset, to ensure no overlap with the training set.  This is a very strong condition in Fourier space,  guaranteeing that any amplitude of a particular mode cannot be generated in both the training and testing sets. Therefore, the pipeline cannot perform well in testing by simply learning a Fourier representation of the training tracks. Even if transformed to another basis, these spaces will still be disjoint, and crucially, the network would be unable to succeed by learning \textit{any} particular basis representation of tracks. The tracker's learning must be basis-independent for successful reconstruction of tracks in a testing set disjoint from its training set. To succeed, the pipeline must learn more general properties of smoothness. Figure~
\ref{fig:manysf} gives a visual representation of  disjoint spaces. 
Examples from the disjoint training set 6-5 are shown in Fig.~\ref{fig:6-5}.

\begin{figure}[h]
    \centering
    \includegraphics[width=0.75\linewidth]{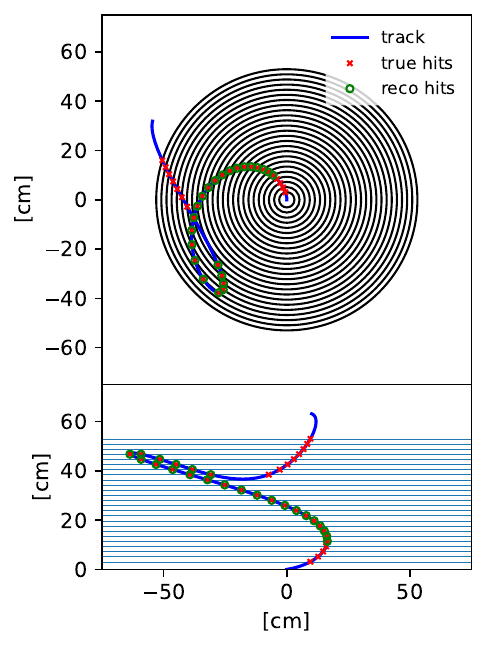}
    \includegraphics[width=0.75\linewidth]{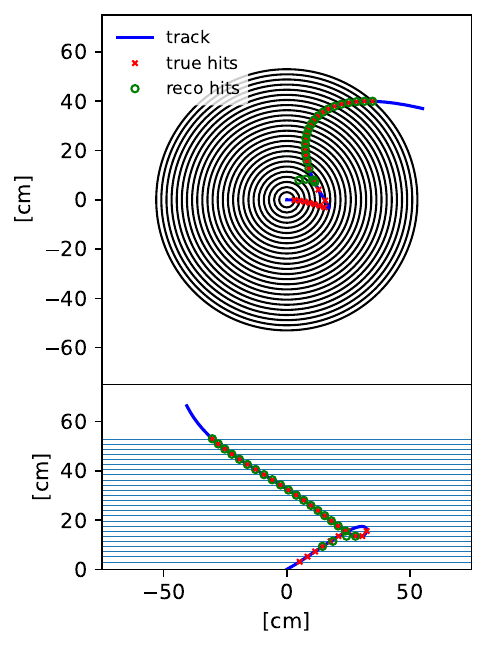}
    \caption{Examples of tracks generated in the 4 - 3 space. The reconstructed (predicted) tracks are shown as well.}
    \label{fig:tracks-43}
\end{figure}


Training and testing samples again consist of $pp \rightarrow t \bar{t}$ events  with a single added non-helical track. In every case, the training sample contains 40k events and the testing sample 4k events. The majority of these tests result in high efficiency, as shown in Tab.~\ref{tab:generalization-test}, demonstrating generalization beyond the training set.  The disjoint testing sets indicate that the network does not learn a basis-dependent representation of tracks, but learns something more general about smooth tracks, and should therefore be able to generalize its training further to other spaces of smooth tracks.     

In one case, training on Schwartz space 5 and testing on 6-5, gives poor performance. The testing set has very long tracks, as shown in Figure.~\ref{fig:6-5}, which makes reconstruction under the double matching criterion very difficult. Generalization is possible, but not universal or guaranteed.

\subsection{Dependence on Training Functions}
\label{subsec:analysis-train-func}

The selection of the Schwartz function clearly impacts the capacity of the network to find tracks, both within and beyond its training set. The form $f(x)= A e^{-\alpha (x-c)^2}$ is chosen, where $A$ is the maximum amplitude, $\alpha$ represents the width, and $c$ is the location of the center. 

In this section, the dependence of the performance on the values of these parameters for the training function is explored; the values of the testing parameters are fixed at $A =  45,\; \alpha =0.02 , \; \text{and}\; c= 3.5$. These values are chosen  because they each lie in the center of the ranges used for each parameter, making for straightforward comparison of training parameters that are either above or below the testing values.


The $n$th Fourier mode has an amplitude drawn uniformly from $[0, f(n)]$.  A maximum amplitude larger than the detector radius is likely to take the track outside of the detector, preventing oscillatory motion that helps distinguish these tracks from the SM background.  A range of amplitudes between 30 and 60cm is therefore considered, for a detector with radius of 50cm. Efficiencies are somewhat higher for smaller values, though fake rates also increase; see Table \ref{tab:param}. To balance these, $A=52.5$cm is selected.

\begin{table}[h]
    \centering
    \caption{ Dependence of efficiency on maximum amplitude parameter of training function. Efficiency is for non-helical tracks; training and testing samples include SM backgrounds.}
    \label{tab:param}
    \begin{tabular}{lll|rr}
    \hline\hline
      Amplitude & Center & Width & Efficiency & Fake Rate\\ 
    \hline
     30 cm &&& 0.718 $\pm$ 0.007 & 0.184\\
     37.5 cm &&& 0.594 $\pm$ 0.008 & 0.001\\
     45 cm &&& 0.771 $\pm$ 0.007 & 0.186\\
     52.5 cm &&& 0.562 $\pm$ 0.008 & 0.001\\
     60 cm &&& 0.562 $\pm$ 0.008 & 0.001\\
     \hline
       &0 && 0.562 $\pm$ 0.008 & 0.001\\
     &1.75 && 0.553 $\pm$ 0.008 & 0.001\\
     &3.5 && 0.546 $\pm$ 0.008 & 0.001\\ 
    &5.25 && 0.546 $\pm$ 0.008 & 0.001\\
    &7 && 0.536 $\pm$ 0.008 & 0.001\\
     \hline
     &&0.01 & 0.562 $\pm$ 0.008 & 0.001\\
     &&0.015 & 0.562 $\pm$ 0.008 & 0.001\\
    &&0.02 & 0.568 $\pm$ 0.008 & 0.001\\
     &&0.025 & 0.594 $\pm$ 0.008 & 0.001\\
    &&0.03 & 0.581 $\pm$ 0.008 & 0.001\\
    &&0.035 & 0.577 $\pm$ 0.008 & 0.001\\
    \hline\hline
    \end{tabular}
    
\end{table}

A similar analysis is performed for the center and width of the training functions. The position of the center is varied in Table \ref{tab:param}. The efficiency decreases monotonically as the center shifts farther from zero. A center at the origin is chosen.

Table \ref{tab:param} shows the efficiency under variation of the width parameter.  
A local maximum of efficiency is reached  at a width of 0.025.   The final training function corresponds to Schwartz function 19 in Tab~\ref{tab:sf}.

\subsection{Results}
\label{results}
The efficiency of the pipeline when trained using Schwartz function 19 is shown in Tab. \ref{tab:final-results} for a selection of testing samples. The training sample size has been increased to 65K events, each with one non-helical track and SM tracks from one SM event. This pipeline has a strong capacity to reconstruct many varieties of smooth tracks when correctly trained, while maintaining a low fake rate.

\begin{table}[h]
    \centering
    \caption{Tracking efficiency and fake rate of a pipeline trained on SM and Schwartz Set 19, for several testing samples.}
        \label{tab:final-results}
    \begin{tabular}{l|rr}
    \hline\hline
    Test & Efficiency & Fake Rate\\ 
    \hline
    SM + Schwartz Set 1 & 0.764 $\pm$ 0.007 & 0.000 \\
    SM + Schwartz Set 2 & 0.755 $\pm$ 0.007 &0.000
    \\
    SM + Schwartz Set 4 & 0.489 $\pm$ 0.008 & 0.001 \\
    SM + Schwartz Set 7 & 0.472 $\pm$ 0.008 & 0.001 \\
    SM + Schwartz Set 8 & 0.662 $\pm$ 0.008 & 0.001 \\
    SM + Schwartz Set 9 & 0.627 $\pm$ 0.008 & 0.001 \\
    SM + Schwartz Set 10 & 0.491 $\pm$ 0.008 & 0.001 \\
    SM + Schwartz Set 19 & 0.504 $\pm$ 0.008 & 0.001 \\
     \hline
    SM + Schwartz Set 21-19 & 0.647 $\pm$ 0.010 & 0.001 \\
    SM + Schwartz Set 22-19 & 0.590 $\pm$ 0.010 & 0.001 \\
    SM + Schwartz Set 23-19 & 0.372 $\pm$ 0.010 & 0.001 \\
    \hline\hline
    \end{tabular}
\end{table}
The ability of the pipeline to generalize when trained on Schwartz function 19 is also shown, by the performance reported in Tab. \ref{tab:final-results} when testing on several spaces that are disjoint from set 19. Performance remains high, thus it is expected that the pipeline has a strong ability to generalize when trained on this function.

\subsection{Quirk Reconstruction}
The non-helical track finding pipeline is applied in this Section to a specific BSM theory, quirks \cite{Kang:2008ea} which leave non-helical tracks. A measure of the pipeline's ability to recover physically motivated signals is provided by this Section.



\begin{figure}[H]

    \centering

    \includegraphics[width=0.70\linewidth]{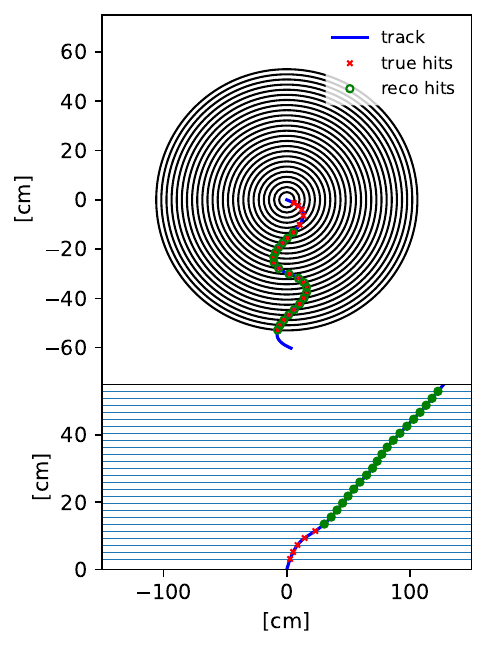}
    \includegraphics[width=0.70\linewidth]{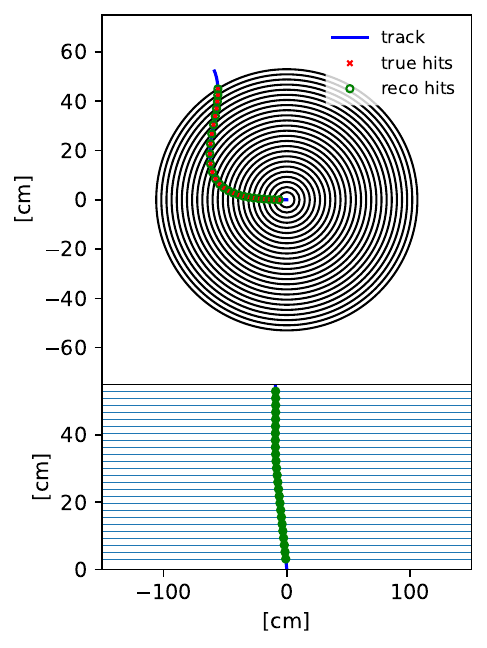}

    \caption{Examples of reconstructed (predicted) quirk tracks}
    \label{fig:quirk_plot}

\end{figure}

\begin{figure}[H]

    \centering

    \includegraphics[scale = 0.45]{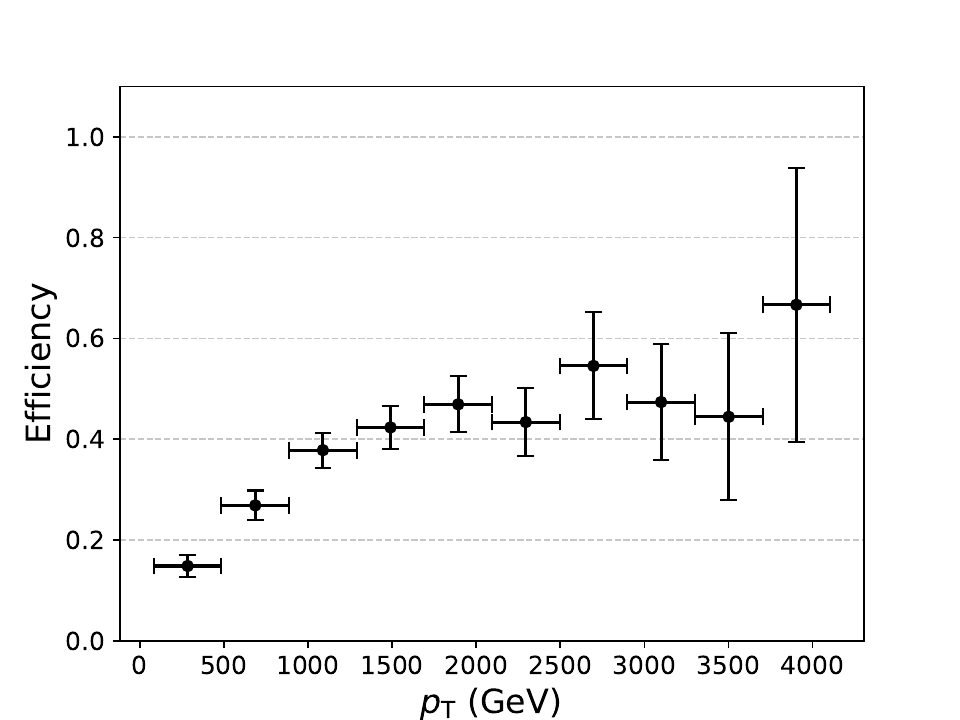}
    \includegraphics[scale = 0.45]{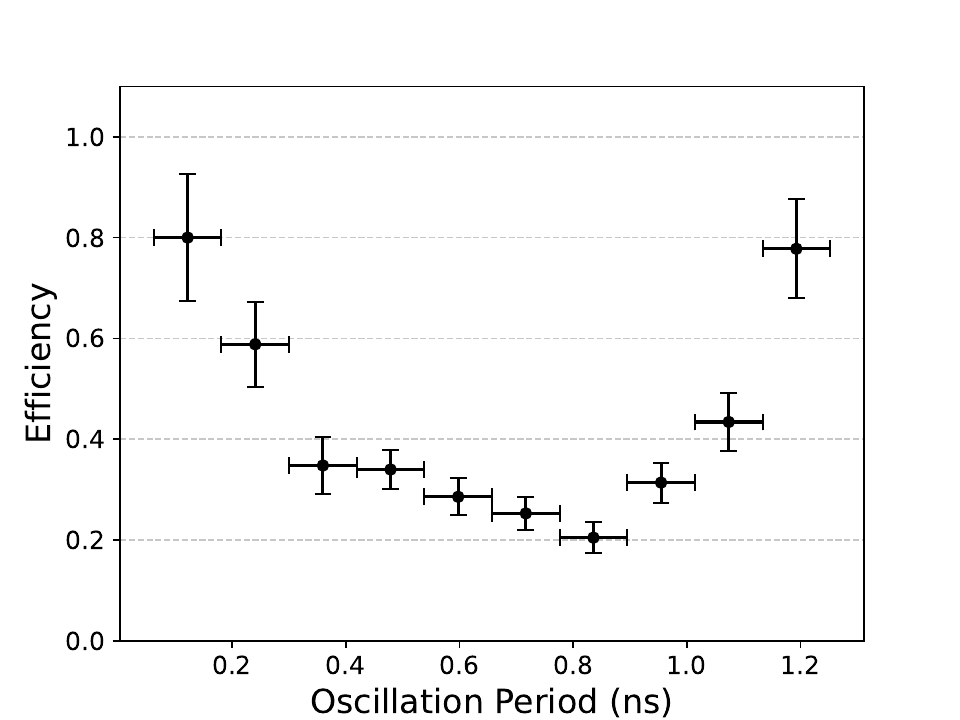}
    \includegraphics[scale = 0.45]{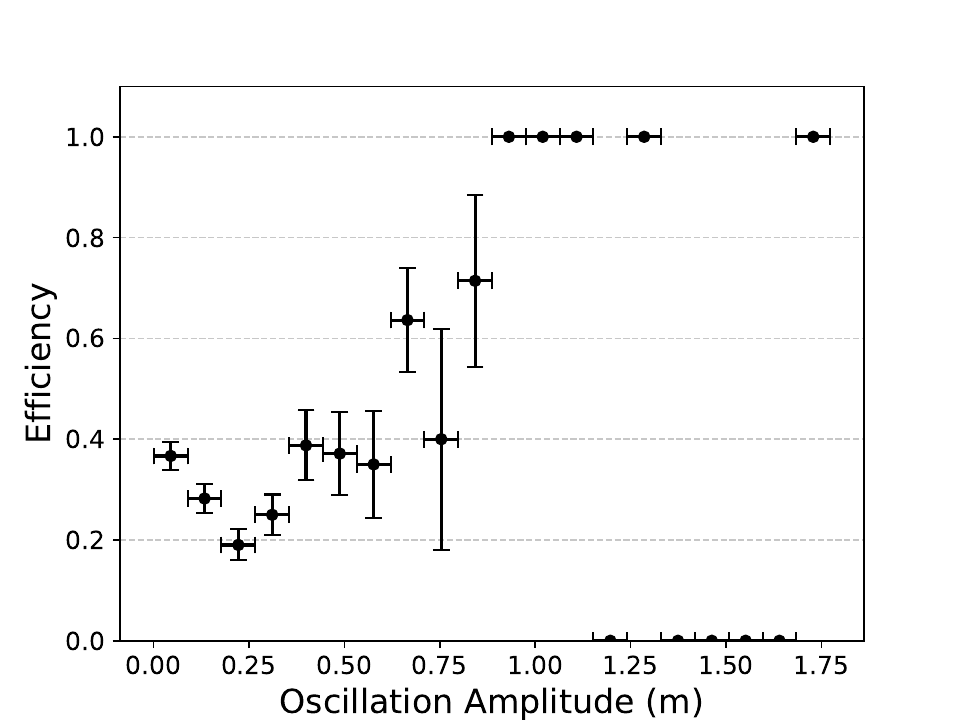}
    \includegraphics[scale = 0.45]{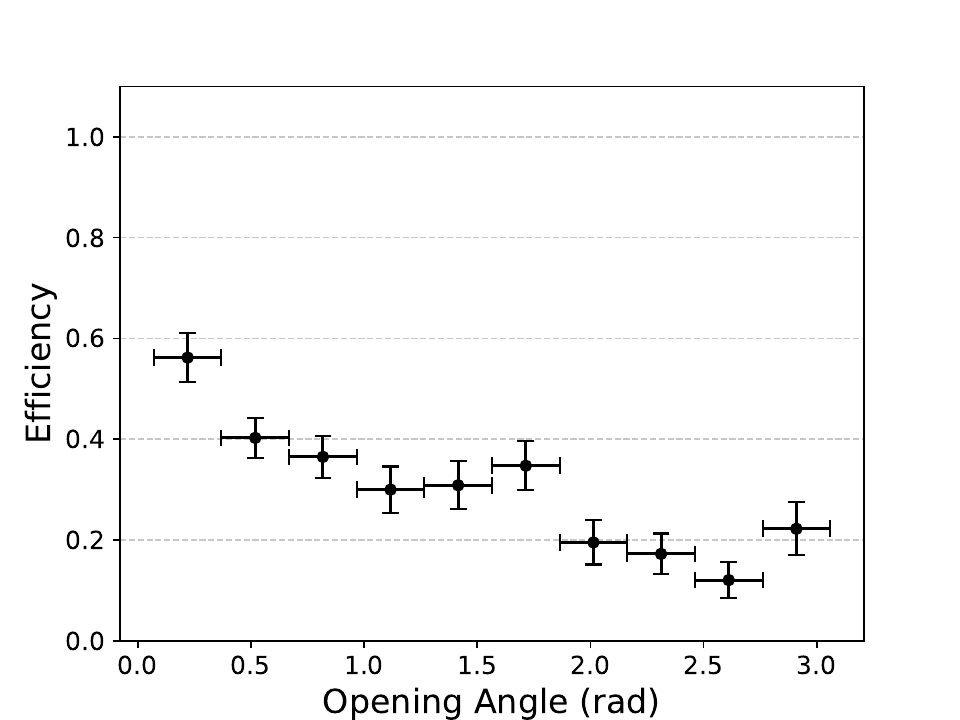}
    \caption{Shown is the reconstruction efficiency on quirks with respect to $p_\mathrm{T}$, oscillation period, amplitude, and opening angle}
    \label{fig:quirk_eff}

\end{figure}

Training is again performed on 65K events with a combination of non-helical tracks from Schwartz function 19 and SM tracks as in Section \ref{results}, and testing is performed on 1K events containing both SM and quirk tracks. Following the quirk generation scheme of \cite{Sha:2024hzq}, an overall efficiency of 32 percent on quirk tracks with 0 fakes is reported. The reconstruction efficiency versus the $p_\mathrm{T}$ of the quirk pair, the opening angle between pair produced quirks, the oscillation amplitude, and the oscillation period is displayed in Figure \ref{fig:quirk_eff}. Examples of reconstructed quirk tracks are shown in Figure ~\ref{fig:quirk_plot}.

\subsection{Rejection of Standard Model tracks}

 To claim the discovery of a new particle from a small number of non-helical tracks requires some technique to distinguish them from the large background of SM tracks, which are also reconstructed by the pipeline.

\begin{figure}[h!]
    \centering
    \includegraphics[width=0.95\linewidth]{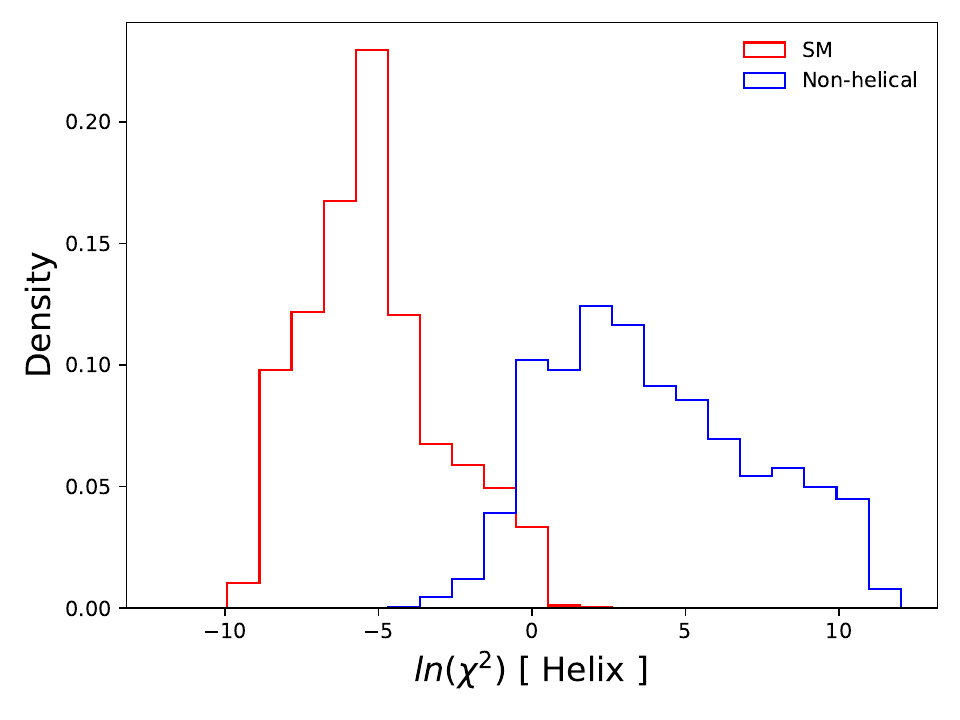}
    \caption{ Distribution of $\chi^2$ values for helical fits to helical SM and non-helical tracks.}
    \label{fig:chisq}
\end{figure}

\begin{figure}[h!]
    \centering
    \includegraphics[width=0.95\linewidth]{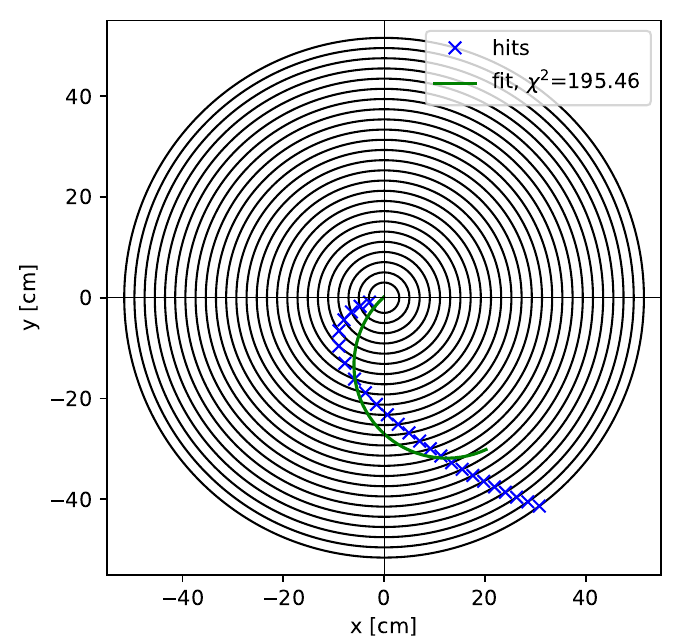}

    \caption{Example helical fit to a non-helical track.}
    \label{fig:fitweird}
\end{figure}

 Helical tracks can be rejected by fitting their points to a helical trajectory and evaluating the goodness of fit, via the $\chi^2$. For each track, an initial rapid estimation of the helical parameters is made~\cite{HANSROUL1988498}, followed by an exploration of parameters values in the vicinity, and refinement with the Nelder-Mead method~\cite{nelder} implemented in scipy.minimize~\cite{2020SciPy-NMeth}.  The distribution of the $\chi^2$ for SM and non-helical tracks is  shown in Figure ~\ref{fig:chisq}.  While some SM tracks will have larger $\chi^2$, and some non-helical tracks have low $\chi^2$, there is a significant population of non-helical tracks with large $\chi^2$ and which are striking by eye. Several examples of fitting helices to hits left by non-helical tracks are shown in Figure .~\ref{fig:fitweird}. 
 
\section{Discussion}
\label{sec:discussion}

The methods presented above can, in principle, generate all possible smooth tracks. In practice, requirements that generation time be finite limit the space somewhat, but still encompass a great variety of trajectories beyond the traditional helix.

Non-helical tracks are extremely challenging for helical tracking codes, as expected. While some traditional algorithms can be modified for specific alternative trajectories, the machine learning pipeline described here has the capacity to learn a broad set of trajectories without the need for explicit parametric specification. The trajectories are specified implicitly via the training sample, and the power of ML to interpolate provides generalization power. Reconstruction performance for non-helical tracks, among a background of helical tracks, is strong, even for tracks outside of the training sample. 

A specific training set is identified which maximizes performance on tracks within that set while maintaining reasonable performance in tracks outside the set.  The fake rate remains very low from hits not associated with real tracks. 

Helical track fitting is a powerful discriminant, but the challenge of filtering helical tracks from non-helical remains significant due to the expected low relative rate.

\section{Conclusions}

The advent of machine learning tools allows for the relaxation of simplifying assumptions in particle tracking that have long limited the power to discover unusual trajectories.  

In a simplified setting, this study provides a proof of principle that non-helical tracks can be found without the need to specify any particular trajectory.  This is a first step towards development of a tracking pipeline which could, for the first time, reveal non-helical surprises that may be lurking in current datasamples and be striking to the naked eye.
\newline
\indent Future work includes more realistic samples, including hit noise models, misalignments, and hard radiation among other effects. The use of physics-oriented constraints that restrict tracks beyond the smoothness assumption, and an investigation of the effects of the Standard Model background process on the performance of this pipeline should also be studied. Other work could include hybrid approaches where the helical tracks are fitted and removed, and the remaining hits are fitted to model agnostic tracks following the methods in this work. Additionally, further study of detector design could explore the capacity to which recording additional information with each hit, such as directional and timing information, could enhance discovery power when combined with works such as this. 

\section*{Acknowledgements}
The authors are grateful to Makayla Vessella and Paata Ivanisvili for helpful discussion. LC and DW are supported by the DOE Office of Science.

\nocite{*}
\bibliographystyle{plainnat}  


\begin{thebibliography}{56}
\expandafter\ifx\csname natexlab\endcsname\relax\def\natexlab#1{#1}\fi
\expandafter\ifx\csname bibnamefont\endcsname\relax
  \def\bibnamefont#1{#1}\fi
\expandafter\ifx\csname bibfnamefont\endcsname\relax
  \def\bibfnamefont#1{#1}\fi
\expandafter\ifx\csname citenamefont\endcsname\relax
  \def\citenamefont#1{#1}\fi
\expandafter\ifx\csname url\endcsname\relax
  \def\url#1{\texttt{#1}}\fi
\expandafter\ifx\csname urlprefix\endcsname\relax\def\urlprefix{URL }\fi
\providecommand{\bibinfo}[2]{#2}
\providecommand{\eprint}[2][]{\url{#2}}

\bibitem[{\citenamefont{Aad et~al.}(2012)}]{ATLAS:2012yve}
\bibinfo{author}{\bibfnamefont{G.}~\bibnamefont{Aad}} \bibnamefont{et~al.} (\bibinfo{collaboration}{ATLAS}), \bibinfo{journal}{Phys. Lett. B} \textbf{\bibinfo{volume}{716}}, \bibinfo{pages}{1} (\bibinfo{year}{2012}), \eprint{1207.7214}.

\bibitem[{\citenamefont{Chatrchyan et~al.}(2012)}]{CMS:2012qbp}
\bibinfo{author}{\bibfnamefont{S.}~\bibnamefont{Chatrchyan}} \bibnamefont{et~al.} (\bibinfo{collaboration}{CMS}), \bibinfo{journal}{Phys. Lett. B} \textbf{\bibinfo{volume}{716}}, \bibinfo{pages}{30} (\bibinfo{year}{2012}), \eprint{1207.7235}.

\bibitem[{\citenamefont{Gehrmann and Malaescu}(2022)}]{annurev:/content/journals/10.1146/annurev-nucl-101920-014923}
\bibinfo{author}{\bibfnamefont{T.}~\bibnamefont{Gehrmann}} \bibnamefont{and} \bibinfo{author}{\bibfnamefont{B.}~\bibnamefont{Malaescu}}, \bibinfo{journal}{Annual Review of Nuclear and Particle Science} \textbf{\bibinfo{volume}{72}}, \bibinfo{pages}{233} (\bibinfo{year}{2022}), ISSN \bibinfo{issn}{1545-4134}.

\bibitem[{\citenamefont{Ferreira~da Silva}(2023)}]{annurev:/content/journals/10.1146/annurev-nucl-102419-052854}
\bibinfo{author}{\bibfnamefont{P.}~\bibnamefont{Ferreira~da Silva}}, \bibinfo{journal}{Annual Review of Nuclear and Particle Science} \textbf{\bibinfo{volume}{73}}, \bibinfo{pages}{255} (\bibinfo{year}{2023}), ISSN \bibinfo{issn}{1545-4134}.

\bibitem[{\citenamefont{Aaij et~al.}(2015)}]{LHCb:2015yax}
\bibinfo{author}{\bibfnamefont{R.}~\bibnamefont{Aaij}} \bibnamefont{et~al.} (\bibinfo{collaboration}{LHCb}), \bibinfo{journal}{Phys. Rev. Lett.} \textbf{\bibinfo{volume}{115}}, \bibinfo{pages}{072001} (\bibinfo{year}{2015}), \eprint{1507.03414}.

\bibitem[{\citenamefont{Aaij et~al.}(2014)}]{LHCb:2014vgu}
\bibinfo{author}{\bibfnamefont{R.}~\bibnamefont{Aaij}} \bibnamefont{et~al.} (\bibinfo{collaboration}{LHCb}), \bibinfo{journal}{Phys. Rev. Lett.} \textbf{\bibinfo{volume}{113}}, \bibinfo{pages}{151601} (\bibinfo{year}{2014}), \eprint{1406.6482}.

\bibitem[{\citenamefont{Aad et~al.}(2010)}]{ATLAS:2010isq}
\bibinfo{author}{\bibfnamefont{G.}~\bibnamefont{Aad}} \bibnamefont{et~al.} (\bibinfo{collaboration}{ATLAS}), \bibinfo{journal}{Phys. Rev. Lett.} \textbf{\bibinfo{volume}{105}}, \bibinfo{pages}{252303} (\bibinfo{year}{2010}), \eprint{1011.6182}.

\bibitem[{\citenamefont{Khachatryan et~al.}(2010)}]{CMS:2010ifv}
\bibinfo{author}{\bibfnamefont{V.}~\bibnamefont{Khachatryan}} \bibnamefont{et~al.} (\bibinfo{collaboration}{CMS}), \bibinfo{journal}{JHEP} \textbf{\bibinfo{volume}{09}}, \bibinfo{pages}{091} (\bibinfo{year}{2010}), \eprint{1009.4122}.

\bibitem[{\citenamefont{Strandlie and Fruhwirth}(2010)}]{Strandlie:2010zz}
\bibinfo{author}{\bibfnamefont{A.}~\bibnamefont{Strandlie}} \bibnamefont{and} \bibinfo{author}{\bibfnamefont{R.}~\bibnamefont{Fruhwirth}}, \bibinfo{journal}{Rev. Mod. Phys.} \textbf{\bibinfo{volume}{82}}, \bibinfo{pages}{1419} (\bibinfo{year}{2010}).

\bibitem[{\citenamefont{Aad et~al.}(2024)}]{ATLAS:2023iat}
\bibinfo{author}{\bibfnamefont{G.}~\bibnamefont{Aad}} \bibnamefont{et~al.} (\bibinfo{collaboration}{ATLAS}), \bibinfo{journal}{Comput. Softw. Big Sci.} \textbf{\bibinfo{volume}{8}}, \bibinfo{pages}{9} (\bibinfo{year}{2024}), \eprint{2308.09471}.

\bibitem[{\citenamefont{Collaboration}(2020)}]{ATLAS:2019wkg}
\bibinfo{author}{\bibfnamefont{A.}~\bibnamefont{Collaboration}} (\bibinfo{collaboration}{ATLAS}), \bibinfo{journal}{Phys. Rev. Lett.} \textbf{\bibinfo{volume}{124}}, \bibinfo{pages}{031802} (\bibinfo{year}{2020}), \eprint{1905.10130}.

\bibitem[{\citenamefont{Abulencia et~al.}(2006)}]{CDF:2005cvf}
\bibinfo{author}{\bibfnamefont{A.}~\bibnamefont{Abulencia}} \bibnamefont{et~al.} (\bibinfo{collaboration}{CDF}), \bibinfo{journal}{Phys. Rev. Lett.} \textbf{\bibinfo{volume}{96}}, \bibinfo{pages}{201801} (\bibinfo{year}{2006}), \eprint{hep-ex/0509015}.

\bibitem[{\citenamefont{Kang and Luty}(2009)}]{Kang:2008ea}
\bibinfo{author}{\bibfnamefont{J.}~\bibnamefont{Kang}} \bibnamefont{and} \bibinfo{author}{\bibfnamefont{M.~A.} \bibnamefont{Luty}}, \bibinfo{journal}{JHEP} \textbf{\bibinfo{volume}{11}}, \bibinfo{pages}{065} (\bibinfo{year}{2009}), \eprint{0805.4642}.

\bibitem[{\citenamefont{Evans and Luty}(2019)}]{Evans:2018jmd}
\bibinfo{author}{\bibfnamefont{J.~A.} \bibnamefont{Evans}} \bibnamefont{and} \bibinfo{author}{\bibfnamefont{M.~A.} \bibnamefont{Luty}}, \bibinfo{journal}{JHEP} \textbf{\bibinfo{volume}{06}}, \bibinfo{pages}{090} (\bibinfo{year}{2019}), \eprint{1811.08903}.

\bibitem[{\citenamefont{Knapen et~al.}(2017)\citenamefont{Knapen, Lou, Papucci, and Setford}}]{Knapen_2017}
\bibinfo{author}{\bibfnamefont{S.}~\bibnamefont{Knapen}}, \bibinfo{author}{\bibfnamefont{H.~K.} \bibnamefont{Lou}}, \bibinfo{author}{\bibfnamefont{M.}~\bibnamefont{Papucci}}, \bibnamefont{and} \bibinfo{author}{\bibfnamefont{J.}~\bibnamefont{Setford}}, \bibinfo{journal}{Physical Review D} \textbf{\bibinfo{volume}{96}} (\bibinfo{year}{2017}), ISSN \bibinfo{issn}{2470-0029}, \urlprefix\url{http://dx.doi.org/10.1103/PhysRevD.96.115015}.

\bibitem[{\citenamefont{Sha et~al.}(2024)\citenamefont{Sha, Murnane, Fieg, Tong, Zakharyan, Fang, and Whiteson}}]{Sha:2024hzq}
\bibinfo{author}{\bibfnamefont{Q.}~\bibnamefont{Sha}}, \bibinfo{author}{\bibfnamefont{D.}~\bibnamefont{Murnane}}, \bibinfo{author}{\bibfnamefont{M.}~\bibnamefont{Fieg}}, \bibinfo{author}{\bibfnamefont{S.}~\bibnamefont{Tong}}, \bibinfo{author}{\bibfnamefont{M.}~\bibnamefont{Zakharyan}}, \bibinfo{author}{\bibfnamefont{Y.}~\bibnamefont{Fang}}, \bibnamefont{and} \bibinfo{author}{\bibfnamefont{D.}~\bibnamefont{Whiteson}} (\bibinfo{year}{2024}), \eprint{2410.00269}.

\bibitem[{\citenamefont{Anderson}(1933)}]{PhysRev.43.491}
\bibinfo{author}{\bibfnamefont{C.~D.} \bibnamefont{Anderson}}, \bibinfo{journal}{Phys. Rev.} \textbf{\bibinfo{volume}{43}}, \bibinfo{pages}{491} (\bibinfo{year}{1933}), \urlprefix\url{https://link.aps.org/doi/10.1103/PhysRev.43.491}.

\bibitem[{\citenamefont{Collaboration}(2017)}]{ATLAS:2017kyn}
\bibinfo{author}{\bibfnamefont{A.}~\bibnamefont{Collaboration}} (\bibinfo{collaboration}{ATLAS}), \bibinfo{journal}{Eur. Phys. J. C} \textbf{\bibinfo{volume}{77}}, \bibinfo{pages}{673} (\bibinfo{year}{2017}), \eprint{1704.07983}.

\bibitem[{\citenamefont{Bronstein et~al.}(2017)\citenamefont{Bronstein, Bruna, LeCun, Szlam, and Vandergheynst}}]{Bronstein:2016thv}
\bibinfo{author}{\bibfnamefont{M.~M.} \bibnamefont{Bronstein}}, \bibinfo{author}{\bibfnamefont{J.}~\bibnamefont{Bruna}}, \bibinfo{author}{\bibfnamefont{Y.}~\bibnamefont{LeCun}}, \bibinfo{author}{\bibfnamefont{A.}~\bibnamefont{Szlam}}, \bibnamefont{and} \bibinfo{author}{\bibfnamefont{P.}~\bibnamefont{Vandergheynst}}, \bibinfo{journal}{IEEE Sig. Proc. Mag.} \textbf{\bibinfo{volume}{34}}, \bibinfo{pages}{18} (\bibinfo{year}{2017}), \eprint{1611.08097}.

\bibitem[{\citenamefont{Ju et~al.}(2020{\natexlab{a}})}]{ExaTrkX:2020nyf}
\bibinfo{author}{\bibfnamefont{X.}~\bibnamefont{Ju}} \bibnamefont{et~al.} (\bibinfo{collaboration}{Exa.TrkX}), in \emph{\bibinfo{booktitle}{{33rd Annual Conference on Neural Information Processing Systems}}} (\bibinfo{year}{2020}{\natexlab{a}}), \eprint{2003.11603}.

\bibitem[{\citenamefont{Ju et~al.}(2021{\natexlab{a}})\citenamefont{Ju, Murnane, Calafiura, Choma, Conlon, Farrell, Xu, Spiropulu, Vlimant, Aurisano et~al.}}]{Ju_2021}
\bibinfo{author}{\bibfnamefont{X.}~\bibnamefont{Ju}}, \bibinfo{author}{\bibfnamefont{D.}~\bibnamefont{Murnane}}, \bibinfo{author}{\bibfnamefont{P.}~\bibnamefont{Calafiura}}, \bibinfo{author}{\bibfnamefont{N.}~\bibnamefont{Choma}}, \bibinfo{author}{\bibfnamefont{S.}~\bibnamefont{Conlon}}, \bibinfo{author}{\bibfnamefont{S.}~\bibnamefont{Farrell}}, \bibinfo{author}{\bibfnamefont{Y.}~\bibnamefont{Xu}}, \bibinfo{author}{\bibfnamefont{M.}~\bibnamefont{Spiropulu}}, \bibinfo{author}{\bibfnamefont{J.-R.} \bibnamefont{Vlimant}}, \bibinfo{author}{\bibfnamefont{A.}~\bibnamefont{Aurisano}}, \bibnamefont{et~al.}, \bibinfo{journal}{The European Physical Journal C} \textbf{\bibinfo{volume}{81}} (\bibinfo{year}{2021}{\natexlab{a}}), ISSN \bibinfo{issn}{1434-6052}, \urlprefix\url{http://dx.doi.org/10.1140/epjc/s10052-021-09675-8}.

\bibitem[{\citenamefont{Caillou et~al.}(2024)\citenamefont{Caillou, Calafiura, Ju, Murnane, Pham, Rougier, Stark, and Vallier}}]{Caillou:2024smf}
\bibinfo{author}{\bibfnamefont{S.}~\bibnamefont{Caillou}}, \bibinfo{author}{\bibfnamefont{P.}~\bibnamefont{Calafiura}}, \bibinfo{author}{\bibfnamefont{X.}~\bibnamefont{Ju}}, \bibinfo{author}{\bibfnamefont{D.}~\bibnamefont{Murnane}}, \bibinfo{author}{\bibfnamefont{T.}~\bibnamefont{Pham}}, \bibinfo{author}{\bibfnamefont{C.}~\bibnamefont{Rougier}}, \bibinfo{author}{\bibfnamefont{J.}~\bibnamefont{Stark}}, \bibnamefont{and} \bibinfo{author}{\bibfnamefont{A.}~\bibnamefont{Vallier}} (\bibinfo{collaboration}{ATLAS}), \bibinfo{journal}{EPJ Web Conf.} \textbf{\bibinfo{volume}{295}}, \bibinfo{pages}{03030} (\bibinfo{year}{2024}).

\bibitem[{\citenamefont{Caillou et~al.}(2022{\natexlab{a}})\citenamefont{Caillou, Calafiura, Farrell, Ju, Murnane, Rougier, Stark, and Vallier}}]{Caillou:2022hly}
\bibinfo{author}{\bibfnamefont{S.}~\bibnamefont{Caillou}}, \bibinfo{author}{\bibfnamefont{P.}~\bibnamefont{Calafiura}}, \bibinfo{author}{\bibfnamefont{S.~A.} \bibnamefont{Farrell}}, \bibinfo{author}{\bibfnamefont{X.}~\bibnamefont{Ju}}, \bibinfo{author}{\bibfnamefont{D.~T.} \bibnamefont{Murnane}}, \bibinfo{author}{\bibfnamefont{C.}~\bibnamefont{Rougier}}, \bibinfo{author}{\bibfnamefont{J.}~\bibnamefont{Stark}}, \bibnamefont{and} \bibinfo{author}{\bibfnamefont{A.}~\bibnamefont{Vallier}} (\bibinfo{collaboration}{ATLAS}) (\bibinfo{year}{2022}{\natexlab{a}}).

\bibitem[{\citenamefont{Alwall et~al.}(2014)\citenamefont{Alwall, Frederix, Frixione, Hirschi, Maltoni, Mattelaer, Shao, Stelzer, Torrielli, and Zaro}}]{Alwall:2014hca}
\bibinfo{author}{\bibfnamefont{J.}~\bibnamefont{Alwall}}, \bibinfo{author}{\bibfnamefont{R.}~\bibnamefont{Frederix}}, \bibinfo{author}{\bibfnamefont{S.}~\bibnamefont{Frixione}}, \bibinfo{author}{\bibfnamefont{V.}~\bibnamefont{Hirschi}}, \bibinfo{author}{\bibfnamefont{F.}~\bibnamefont{Maltoni}}, \bibinfo{author}{\bibfnamefont{O.}~\bibnamefont{Mattelaer}}, \bibinfo{author}{\bibfnamefont{H.~S.} \bibnamefont{Shao}}, \bibinfo{author}{\bibfnamefont{T.}~\bibnamefont{Stelzer}}, \bibinfo{author}{\bibfnamefont{P.}~\bibnamefont{Torrielli}}, \bibnamefont{and} \bibinfo{author}{\bibfnamefont{M.}~\bibnamefont{Zaro}}, \bibinfo{journal}{JHEP} \textbf{\bibinfo{volume}{07}}, \bibinfo{pages}{079} (\bibinfo{year}{2014}), \eprint{1405.0301}.

\bibitem[{\citenamefont{Bierlich et~al.}(2022)}]{Bierlich:2022pfr}
\bibinfo{author}{\bibfnamefont{C.}~\bibnamefont{Bierlich}} \bibnamefont{et~al.}, \bibinfo{journal}{SciPost Phys. Codeb.} \textbf{\bibinfo{volume}{2022}}, \bibinfo{pages}{8} (\bibinfo{year}{2022}), \eprint{2203.11601}.

\bibitem[{\citenamefont{Ju et~al.}(2021{\natexlab{b}})}]{ExaTrkX:2021abe}
\bibinfo{author}{\bibfnamefont{X.}~\bibnamefont{Ju}} \bibnamefont{et~al.} (\bibinfo{collaboration}{Exa.TrkX}), \bibinfo{journal}{Eur. Phys. J. C} \textbf{\bibinfo{volume}{81}}, \bibinfo{pages}{876} (\bibinfo{year}{2021}{\natexlab{b}}), \eprint{2103.06995}.

\bibitem[{\citenamefont{Landau and Lifschits}(1975)}]{Landau:1975pou}
\bibinfo{author}{\bibfnamefont{L.~D.} \bibnamefont{Landau}} \bibnamefont{and} \bibinfo{author}{\bibfnamefont{E.~M.} \bibnamefont{Lifschits}}, \emph{\bibinfo{title}{{The Classical Theory of Fields}}}, vol. \bibinfo{volume}{Volume 2} of \emph{\bibinfo{series}{Course of Theoretical Physics}} (\bibinfo{publisher}{Pergamon Press}, \bibinfo{address}{Oxford}, \bibinfo{year}{1975}), ISBN \bibinfo{isbn}{978-0-08-018176-9}.

\bibitem[{\citenamefont{Stein}(2003)}]{Stein:2003}
\bibinfo{author}{\bibfnamefont{R.}~\bibnamefont{Stein}, \bibfnamefont{Elias M.;~Shakarchi}}, \emph{\bibinfo{title}{Fourier Analysis: An Introduction (Princeton Lectures in Analysis I)}} (\bibinfo{publisher}{Princeton: Princeton University Press}, \bibinfo{year}{2003}).

\bibitem[{\citenamefont{Hormander}(1990)}]{Hormander:1990}
\bibinfo{author}{\bibnamefont{Hormander}}, \emph{\bibinfo{title}{The Analysis of Linear Partial Differential Operators 1}} (\bibinfo{publisher}{Berlin: Springer-Verlag}, \bibinfo{year}{1990}).

\bibitem[{\citenamefont{Reed}(1980)}]{Reed:1980}
\bibinfo{author}{\bibfnamefont{B.}~\bibnamefont{Reed}, \bibfnamefont{M.;~Simon}}, \emph{\bibinfo{title}{Methods of Modern Mathematical Physics: Functional Analysis I (Revised and enlarged ed.)}} (\bibinfo{publisher}{San Diego: Academic Press}, \bibinfo{year}{1980}).

\bibitem[{\citenamefont{Duarte and Vlimant}(2022)}]{duarte2022graph}
\bibinfo{author}{\bibfnamefont{J.}~\bibnamefont{Duarte}} \bibnamefont{and} \bibinfo{author}{\bibfnamefont{J.-R.} \bibnamefont{Vlimant}}, in \emph{\bibinfo{booktitle}{Artificial intelligence for high energy physics}} (\bibinfo{publisher}{World Scientific}, \bibinfo{year}{2022}), pp. \bibinfo{pages}{387--436}.

\bibitem[{\citenamefont{DeZoort et~al.}(2023)\citenamefont{DeZoort, Battaglia, Biscarat, and Vlimant}}]{dezoort2023graph}
\bibinfo{author}{\bibfnamefont{G.}~\bibnamefont{DeZoort}}, \bibinfo{author}{\bibfnamefont{P.~W.} \bibnamefont{Battaglia}}, \bibinfo{author}{\bibfnamefont{C.}~\bibnamefont{Biscarat}}, \bibnamefont{and} \bibinfo{author}{\bibfnamefont{J.-R.} \bibnamefont{Vlimant}}, \bibinfo{journal}{Nature Reviews Physics} \textbf{\bibinfo{volume}{5}}, \bibinfo{pages}{281} (\bibinfo{year}{2023}).

\bibitem[{\citenamefont{Thais et~al.}(2022)\citenamefont{Thais, Calafiura, Chachamis, DeZoort, Duarte, Ganguly, Kagan, Murnane, Neubauer, and Terao}}]{thais2022graph}
\bibinfo{author}{\bibfnamefont{S.}~\bibnamefont{Thais}}, \bibinfo{author}{\bibfnamefont{P.}~\bibnamefont{Calafiura}}, \bibinfo{author}{\bibfnamefont{G.}~\bibnamefont{Chachamis}}, \bibinfo{author}{\bibfnamefont{G.}~\bibnamefont{DeZoort}}, \bibinfo{author}{\bibfnamefont{J.}~\bibnamefont{Duarte}}, \bibinfo{author}{\bibfnamefont{S.}~\bibnamefont{Ganguly}}, \bibinfo{author}{\bibfnamefont{M.}~\bibnamefont{Kagan}}, \bibinfo{author}{\bibfnamefont{D.}~\bibnamefont{Murnane}}, \bibinfo{author}{\bibfnamefont{M.~S.} \bibnamefont{Neubauer}}, \bibnamefont{and} \bibinfo{author}{\bibfnamefont{K.}~\bibnamefont{Terao}}, \bibinfo{journal}{arXiv preprint arXiv:2203.12852}  (\bibinfo{year}{2022}).

\bibitem[{\citenamefont{Heinrich et~al.}(2024)\citenamefont{Heinrich, Huth, Salzburger, and Wettig}}]{heinrich2024combined}
\bibinfo{author}{\bibfnamefont{L.}~\bibnamefont{Heinrich}}, \bibinfo{author}{\bibfnamefont{B.}~\bibnamefont{Huth}}, \bibinfo{author}{\bibfnamefont{A.}~\bibnamefont{Salzburger}}, \bibnamefont{and} \bibinfo{author}{\bibfnamefont{T.}~\bibnamefont{Wettig}}, \bibinfo{journal}{arXiv preprint arXiv:2401.16016}  (\bibinfo{year}{2024}).

\bibitem[{\citenamefont{Vaage}(2022)}]{vaage2022reinforcement}
\bibinfo{author}{\bibfnamefont{L.~H.} \bibnamefont{Vaage}}, in \emph{\bibinfo{booktitle}{Connecting the Dots Workshop (CTD)}} (\bibinfo{year}{2022}).

\bibitem[{\citenamefont{Kortus et~al.}(2023)\citenamefont{Kortus, Keidel, Gauger, pCT Collaboration et~al.}}]{kortus2023towards}
\bibinfo{author}{\bibfnamefont{T.}~\bibnamefont{Kortus}}, \bibinfo{author}{\bibfnamefont{R.}~\bibnamefont{Keidel}}, \bibinfo{author}{\bibfnamefont{N.~R.} \bibnamefont{Gauger}}, \bibinfo{author}{\bibfnamefont{B.}~\bibnamefont{pCT Collaboration}}, \bibnamefont{et~al.}, \bibinfo{journal}{IEEE Transactions on Pattern Analysis and Machine Intelligence}  (\bibinfo{year}{2023}).

\bibitem[{\citenamefont{Caillou et~al.}(2022{\natexlab{b}})\citenamefont{Caillou, Calafiura, Rougier, Stark, Murnane, Vallier, Ju, and Farrell}}]{caillou2022atlas}
\bibinfo{author}{\bibfnamefont{S.}~\bibnamefont{Caillou}}, \bibinfo{author}{\bibfnamefont{P.}~\bibnamefont{Calafiura}}, \bibinfo{author}{\bibfnamefont{C.}~\bibnamefont{Rougier}}, \bibinfo{author}{\bibfnamefont{J.}~\bibnamefont{Stark}}, \bibinfo{author}{\bibfnamefont{D.~T.} \bibnamefont{Murnane}}, \bibinfo{author}{\bibfnamefont{A.}~\bibnamefont{Vallier}}, \bibinfo{author}{\bibfnamefont{X.}~\bibnamefont{Ju}}, \bibnamefont{and} \bibinfo{author}{\bibfnamefont{S.~A.} \bibnamefont{Farrell}}, \bibinfo{type}{Tech. Rep.}, \bibinfo{institution}{ATL-COM-ITK-2022-057} (\bibinfo{year}{2022}{\natexlab{b}}).

\bibitem[{\citenamefont{Verma and Jena}(2020)}]{verma2020particle}
\bibinfo{author}{\bibfnamefont{Y.}~\bibnamefont{Verma}} \bibnamefont{and} \bibinfo{author}{\bibfnamefont{S.}~\bibnamefont{Jena}}, \bibinfo{journal}{arXiv preprint arXiv:2012.08515}  (\bibinfo{year}{2020}).

\bibitem[{\citenamefont{Hewes et~al.}(2021)\citenamefont{Hewes, Aurisano, Cerati, Kowalkowski, Lee, Liao, Day, Agrawal, Spiropulu, Vlimant et~al.}}]{hewes2021graph}
\bibinfo{author}{\bibfnamefont{J.}~\bibnamefont{Hewes}}, \bibinfo{author}{\bibfnamefont{A.}~\bibnamefont{Aurisano}}, \bibinfo{author}{\bibfnamefont{G.}~\bibnamefont{Cerati}}, \bibinfo{author}{\bibfnamefont{J.}~\bibnamefont{Kowalkowski}}, \bibinfo{author}{\bibfnamefont{C.}~\bibnamefont{Lee}}, \bibinfo{author}{\bibfnamefont{W.-k.} \bibnamefont{Liao}}, \bibinfo{author}{\bibfnamefont{A.}~\bibnamefont{Day}}, \bibinfo{author}{\bibfnamefont{A.}~\bibnamefont{Agrawal}}, \bibinfo{author}{\bibfnamefont{M.}~\bibnamefont{Spiropulu}}, \bibinfo{author}{\bibfnamefont{J.-R.} \bibnamefont{Vlimant}}, \bibnamefont{et~al.}, in \emph{\bibinfo{booktitle}{EPJ Web of Conferences}} (\bibinfo{organization}{EDP Sciences}, \bibinfo{year}{2021}), vol. \bibinfo{volume}{251}, p. \bibinfo{pages}{03054}.

\bibitem[{\citenamefont{Drielsma et~al.}(2021)\citenamefont{Drielsma, Terao, Domin{\'e}, and Koh}}]{drielsma2021scalable}
\bibinfo{author}{\bibfnamefont{F.}~\bibnamefont{Drielsma}}, \bibinfo{author}{\bibfnamefont{K.}~\bibnamefont{Terao}}, \bibinfo{author}{\bibfnamefont{L.}~\bibnamefont{Domin{\'e}}}, \bibnamefont{and} \bibinfo{author}{\bibfnamefont{D.~H.} \bibnamefont{Koh}}, \bibinfo{journal}{arXiv preprint arXiv:2102.01033}  (\bibinfo{year}{2021}).

\bibitem[{\citenamefont{Akram and Ju}(2022)}]{akram2022track}
\bibinfo{author}{\bibfnamefont{A.}~\bibnamefont{Akram}} \bibnamefont{and} \bibinfo{author}{\bibfnamefont{X.}~\bibnamefont{Ju}}, \bibinfo{journal}{arXiv preprint arXiv:2208.12178}  (\bibinfo{year}{2022}).

\bibitem[{\citenamefont{Jia et~al.}(2024)\citenamefont{Jia, Qin, Li, Huang, Zhang, Yin, Zhang, and Yuan}}]{jia2024besiii}
\bibinfo{author}{\bibfnamefont{X.}~\bibnamefont{Jia}}, \bibinfo{author}{\bibfnamefont{X.}~\bibnamefont{Qin}}, \bibinfo{author}{\bibfnamefont{T.}~\bibnamefont{Li}}, \bibinfo{author}{\bibfnamefont{X.}~\bibnamefont{Huang}}, \bibinfo{author}{\bibfnamefont{X.}~\bibnamefont{Zhang}}, \bibinfo{author}{\bibfnamefont{N.}~\bibnamefont{Yin}}, \bibinfo{author}{\bibfnamefont{Y.}~\bibnamefont{Zhang}}, \bibnamefont{and} \bibinfo{author}{\bibfnamefont{Y.}~\bibnamefont{Yuan}}, in \emph{\bibinfo{booktitle}{EPJ Web of Conferences}} (\bibinfo{organization}{EDP Sciences}, \bibinfo{year}{2024}), vol. \bibinfo{volume}{295}, p. \bibinfo{pages}{09006}.

\bibitem[{\citenamefont{Biscarat et~al.}(2021)\citenamefont{Biscarat, Caillou, Rougier, Stark, and Zahreddine}}]{biscarat2021towards}
\bibinfo{author}{\bibfnamefont{C.}~\bibnamefont{Biscarat}}, \bibinfo{author}{\bibfnamefont{S.}~\bibnamefont{Caillou}}, \bibinfo{author}{\bibfnamefont{C.}~\bibnamefont{Rougier}}, \bibinfo{author}{\bibfnamefont{J.}~\bibnamefont{Stark}}, \bibnamefont{and} \bibinfo{author}{\bibfnamefont{J.}~\bibnamefont{Zahreddine}}, in \emph{\bibinfo{booktitle}{EPJ Web of Conferences}} (\bibinfo{organization}{EDP Sciences}, \bibinfo{year}{2021}), vol. \bibinfo{volume}{251}, p. \bibinfo{pages}{03047}.

\bibitem[{\citenamefont{Zdybal et~al.}(2024)\citenamefont{Zdybal, Kucharczyk, and Wolter}}]{zdybal2024machine}
\bibinfo{author}{\bibfnamefont{M.}~\bibnamefont{Zdybal}}, \bibinfo{author}{\bibfnamefont{M.}~\bibnamefont{Kucharczyk}}, \bibnamefont{and} \bibinfo{author}{\bibfnamefont{M.}~\bibnamefont{Wolter}}, \bibinfo{journal}{arXiv preprint arXiv:2402.02913}  (\bibinfo{year}{2024}).

\bibitem[{\citenamefont{Andrews et~al.}(2021)\citenamefont{Andrews, Burkle, Chaudhari, Di~Croce, Gleyzer, Heintz, Narain, Paulini, and Usai}}]{andrews2021accelerating}
\bibinfo{author}{\bibfnamefont{M.}~\bibnamefont{Andrews}}, \bibinfo{author}{\bibfnamefont{B.}~\bibnamefont{Burkle}}, \bibinfo{author}{\bibfnamefont{S.}~\bibnamefont{Chaudhari}}, \bibinfo{author}{\bibfnamefont{D.}~\bibnamefont{Di~Croce}}, \bibinfo{author}{\bibfnamefont{S.}~\bibnamefont{Gleyzer}}, \bibinfo{author}{\bibfnamefont{U.}~\bibnamefont{Heintz}}, \bibinfo{author}{\bibfnamefont{M.}~\bibnamefont{Narain}}, \bibinfo{author}{\bibfnamefont{M.}~\bibnamefont{Paulini}}, \bibnamefont{and} \bibinfo{author}{\bibfnamefont{E.}~\bibnamefont{Usai}}, in \emph{\bibinfo{booktitle}{EPJ Web of Conferences}} (\bibinfo{organization}{EDP Sciences}, \bibinfo{year}{2021}), vol. \bibinfo{volume}{251}, p. \bibinfo{pages}{03057}.

\bibitem[{\citenamefont{DeZoort et~al.}(2021)\citenamefont{DeZoort, Thais, Duarte, Razavimaleki, Atkinson, Ojalvo, Neubauer, and Elmer}}]{dezoort2021charged}
\bibinfo{author}{\bibfnamefont{G.}~\bibnamefont{DeZoort}}, \bibinfo{author}{\bibfnamefont{S.}~\bibnamefont{Thais}}, \bibinfo{author}{\bibfnamefont{J.}~\bibnamefont{Duarte}}, \bibinfo{author}{\bibfnamefont{V.}~\bibnamefont{Razavimaleki}}, \bibinfo{author}{\bibfnamefont{M.}~\bibnamefont{Atkinson}}, \bibinfo{author}{\bibfnamefont{I.}~\bibnamefont{Ojalvo}}, \bibinfo{author}{\bibfnamefont{M.}~\bibnamefont{Neubauer}}, \bibnamefont{and} \bibinfo{author}{\bibfnamefont{P.}~\bibnamefont{Elmer}}, \bibinfo{journal}{Computing and Software for Big Science} \textbf{\bibinfo{volume}{5}}, \bibinfo{pages}{1} (\bibinfo{year}{2021}).

\bibitem[{\citenamefont{Correia et~al.}(2024)\citenamefont{Correia, Giasemis, Garroum, Gligorov, and Granado}}]{Correia_2024}
\bibinfo{author}{\bibfnamefont{A.}~\bibnamefont{Correia}}, \bibinfo{author}{\bibfnamefont{F.~I.} \bibnamefont{Giasemis}}, \bibinfo{author}{\bibfnamefont{N.}~\bibnamefont{Garroum}}, \bibinfo{author}{\bibfnamefont{V.~V.} \bibnamefont{Gligorov}}, \bibnamefont{and} \bibinfo{author}{\bibfnamefont{B.}~\bibnamefont{Granado}}, \bibinfo{journal}{Journal of Instrumentation} \textbf{\bibinfo{volume}{19}}, \bibinfo{pages}{P12022} (\bibinfo{year}{2024}), \urlprefix\url{https://dx.doi.org/10.1088/1748-0221/19/12/P12022}.

\bibitem[{\citenamefont{Elabd et~al.}(2022)\citenamefont{Elabd, Razavimaleki, Huang, Duarte, Atkinson, DeZoort, Elmer, Hauck, Hu, Hsu et~al.}}]{Elabd_2022}
\bibinfo{author}{\bibfnamefont{A.}~\bibnamefont{Elabd}}, \bibinfo{author}{\bibfnamefont{V.}~\bibnamefont{Razavimaleki}}, \bibinfo{author}{\bibfnamefont{S.-Y.} \bibnamefont{Huang}}, \bibinfo{author}{\bibfnamefont{J.}~\bibnamefont{Duarte}}, \bibinfo{author}{\bibfnamefont{M.}~\bibnamefont{Atkinson}}, \bibinfo{author}{\bibfnamefont{G.}~\bibnamefont{DeZoort}}, \bibinfo{author}{\bibfnamefont{P.}~\bibnamefont{Elmer}}, \bibinfo{author}{\bibfnamefont{S.}~\bibnamefont{Hauck}}, \bibinfo{author}{\bibfnamefont{J.-X.} \bibnamefont{Hu}}, \bibinfo{author}{\bibfnamefont{S.-C.} \bibnamefont{Hsu}}, \bibnamefont{et~al.}, \bibinfo{journal}{Frontiers in Big Data} \textbf{\bibinfo{volume}{5}} (\bibinfo{year}{2022}), ISSN \bibinfo{issn}{2624-909X}, \urlprefix\url{http://dx.doi.org/10.3389/fdata.2022.828666}.

\bibitem[{\citenamefont{Hansroul et~al.}(1988)\citenamefont{Hansroul, Jeremie, and Savard}}]{HANSROUL1988498}
\bibinfo{author}{\bibfnamefont{M.}~\bibnamefont{Hansroul}}, \bibinfo{author}{\bibfnamefont{H.}~\bibnamefont{Jeremie}}, \bibnamefont{and} \bibinfo{author}{\bibfnamefont{D.}~\bibnamefont{Savard}}, \bibinfo{journal}{Nuclear Instruments and Methods in Physics Research Section A: Accelerators, Spectrometers, Detectors and Associated Equipment} \textbf{\bibinfo{volume}{270}}, \bibinfo{pages}{498} (\bibinfo{year}{1988}), ISSN \bibinfo{issn}{0168-9002}, \urlprefix\url{https://www.sciencedirect.com/science/article/pii/016890028890722X}.

\bibitem[{\citenamefont{Nelder and Mead}(1965)}]{nelder}
\bibinfo{author}{\bibfnamefont{J.}~\bibnamefont{Nelder}} \bibnamefont{and} \bibinfo{author}{\bibfnamefont{R.}~\bibnamefont{Mead}}, \bibinfo{journal}{Computer Journal} \textbf{\bibinfo{volume}{7}}, \bibinfo{pages}{308} (\bibinfo{year}{1965}).

\bibitem[{\citenamefont{Virtanen et~al.}(2020)\citenamefont{Virtanen, Gommers, Oliphant, Haberland, Reddy, Cournapeau, Burovski, Peterson, Weckesser, Bright et~al.}}]{2020SciPy-NMeth}
\bibinfo{author}{\bibfnamefont{P.}~\bibnamefont{Virtanen}}, \bibinfo{author}{\bibfnamefont{R.}~\bibnamefont{Gommers}}, \bibinfo{author}{\bibfnamefont{T.~E.} \bibnamefont{Oliphant}}, \bibinfo{author}{\bibfnamefont{M.}~\bibnamefont{Haberland}}, \bibinfo{author}{\bibfnamefont{T.}~\bibnamefont{Reddy}}, \bibinfo{author}{\bibfnamefont{D.}~\bibnamefont{Cournapeau}}, \bibinfo{author}{\bibfnamefont{E.}~\bibnamefont{Burovski}}, \bibinfo{author}{\bibfnamefont{P.}~\bibnamefont{Peterson}}, \bibinfo{author}{\bibfnamefont{W.}~\bibnamefont{Weckesser}}, \bibinfo{author}{\bibfnamefont{J.}~\bibnamefont{Bright}}, \bibnamefont{et~al.}, \bibinfo{journal}{Nature Methods} \textbf{\bibinfo{volume}{17}}, \bibinfo{pages}{261} (\bibinfo{year}{2020}).

\bibitem[{\citenamefont{{ATLAS Collaboration}}(2012)}]{Aad_2012}
\bibinfo{author}{\bibnamefont{{ATLAS Collaboration}}}, \bibinfo{journal}{Physics Letters B} \textbf{\bibinfo{volume}{716}} (\bibinfo{year}{2012}), ISSN \bibinfo{issn}{0370-2693}, \urlprefix\url{http://dx.doi.org/10.1016/j.physletb.2012.08.020}.

\bibitem[{\citenamefont{Collaboration}(2012)}]{Chatrchyan_2012}
\bibinfo{author}{\bibfnamefont{C.}~\bibnamefont{Collaboration}}, \bibinfo{journal}{Physics Letters B} \textbf{\bibinfo{volume}{716}} (\bibinfo{year}{2012}), ISSN \bibinfo{issn}{0370-2693}, \urlprefix\url{http://dx.doi.org/10.1016/j.physletb.2012.08.021}.

\bibitem[{\citenamefont{Collaboration}(2016)}]{ATLAS:2015eiz}
\bibinfo{author}{\bibfnamefont{A.}~\bibnamefont{Collaboration}} (\bibinfo{collaboration}{ATLAS}), \bibinfo{journal}{Phys. Rev. D} \textbf{\bibinfo{volume}{93}}, \bibinfo{pages}{052002} (\bibinfo{year}{2016}), \eprint{1509.07152}.

\bibitem[{\citenamefont{Farrell et~al.}(2018)\citenamefont{Farrell, Calafiura, Mudigonda, Prabhat, Anderson, Vlimant, Zheng, Bendavid, Spiropulu, Cerati et~al.}}]{farrell2018novel}
\bibinfo{author}{\bibfnamefont{S.}~\bibnamefont{Farrell}}, \bibinfo{author}{\bibfnamefont{P.}~\bibnamefont{Calafiura}}, \bibinfo{author}{\bibfnamefont{M.}~\bibnamefont{Mudigonda}}, \bibinfo{author}{\bibnamefont{Prabhat}}, \bibinfo{author}{\bibfnamefont{D.}~\bibnamefont{Anderson}}, \bibinfo{author}{\bibfnamefont{J.-R.} \bibnamefont{Vlimant}}, \bibinfo{author}{\bibfnamefont{S.}~\bibnamefont{Zheng}}, \bibinfo{author}{\bibfnamefont{J.}~\bibnamefont{Bendavid}}, \bibinfo{author}{\bibfnamefont{M.}~\bibnamefont{Spiropulu}}, \bibinfo{author}{\bibfnamefont{G.}~\bibnamefont{Cerati}}, \bibnamefont{et~al.}, \emph{\bibinfo{title}{Novel deep learning methods for track reconstruction}} (\bibinfo{year}{2018}), \eprint{1810.06111}.

\bibitem[{\citenamefont{Ju et~al.}(2020{\natexlab{b}})\citenamefont{Ju, Farrell, Calafiura, Murnane, Gray, Klijnsma, Pedro, Cerati, Kowalkowski, Perdue et~al.}}]{ju2020graph}
\bibinfo{author}{\bibfnamefont{X.}~\bibnamefont{Ju}}, \bibinfo{author}{\bibfnamefont{S.}~\bibnamefont{Farrell}}, \bibinfo{author}{\bibfnamefont{P.}~\bibnamefont{Calafiura}}, \bibinfo{author}{\bibfnamefont{D.}~\bibnamefont{Murnane}}, \bibinfo{author}{\bibfnamefont{L.}~\bibnamefont{Gray}}, \bibinfo{author}{\bibfnamefont{T.}~\bibnamefont{Klijnsma}}, \bibinfo{author}{\bibfnamefont{K.}~\bibnamefont{Pedro}}, \bibinfo{author}{\bibfnamefont{G.}~\bibnamefont{Cerati}}, \bibinfo{author}{\bibfnamefont{J.}~\bibnamefont{Kowalkowski}}, \bibinfo{author}{\bibfnamefont{G.}~\bibnamefont{Perdue}}, \bibnamefont{et~al.}, \bibinfo{journal}{arXiv preprint arXiv:2003.11603}  (\bibinfo{year}{2020}{\natexlab{b}}).

\end{thebibliography}

\appendix

\subsection{Schwartz Functions}

\begin{table}[h]
    \centering
       \caption{Selected Schwartz functions.}
    \label{tab:sf}
    \begin{tabular}{c|c}
    \hline    \hline
    Number & Expression \\ 
    \hline
     1 & $f_{1}(n)= 60 \cdot \exp(-0.01 n^2)$ \\
     2 & $f_{2}(n)= 60 \cdot \exp(-0.01 (n-7)^2)$ \\
     3 & $f_{3}(n)= 30 \cdot \exp(-0.01 n^2)$ \\
     4 & $f_{4}(n)= 30 \cdot \exp(-0.01 (n-7)^2)$ \\
     5 & $f_{5}(n)= 10 \cdot \exp(-0.01 n^2)$ \\
     6 & $f_{6}(n)= 10 \cdot \exp(-0.01 (n-7)^2)$ \\
     7 & $f_{7}(n)= 45 \cdot \exp(-0.01 n^2)$ \\
     8 & $f_{8}(n)= 45 \cdot \exp(-0.01 (n-7)^2)$ \\
     9 & $f_{9}(n)= 45 \cdot \exp(-0.01 (n-3.5)^2)$ \\
     10 & $f_{10}(n)= 60 \cdot \exp(-0.03 n^2)$ \\
     11 & $f_{11}(n)= 45 \cdot \exp(-0.02 (n-3.5)^2)$ \\
     12 & $f_{12}(n)= 60 \cdot \exp(-0.01 (n-3.5)^2)$ \\
     13 & $f_{13}(n)= 60 \cdot \exp(-0.02 n^2)$ \\
     14 & $f_{14}(n)= 52.5 \cdot \exp(-0.01 n^2)$ \\
     15 & $f_{15}(n)= 37.5 \cdot \exp(-0.01 n^2)$ \\
     16 & $f_{16}(n)= 60 \cdot \exp(-0.01 (n-1.75)^2)$ \\
     17 & $f_{17}(n)= 60 \cdot \exp(-0.01 (n-5.25)^2)$ \\
     18 & $f_{18}(n)= 60 \cdot \exp(-0.015 n^2)$ \\
     19 & $f_{19}(n)= 60 \cdot \exp(-0.025 n^2)$ \\
     20 & $f_{20}(n)= 60 \cdot \exp(-0.035 n^2)$ \\
     21 & $f_{21}(n)= 60 \cdot \exp(-0.025 (n-3.5)^2)$ \\
     22 & $f_{22}(n)= 45 \cdot \exp(-0.025 (n-7)^2)$ \\
     23 & $f_{23}(n)= 30 \cdot \exp(-0.03 (n-7)^2)$ \\
    \hline    \hline
    \end{tabular}
 
\end{table}

\subsection{Appendix: Schwartz Space Properties}
\label{appendix:shwartz}

It is shown here that the method of smooth track generation detailed in Section \ref{subsec:sig} is capable of generating every smooth track.

Curves are generated by choosing a Schwartz function, $s\in \mathcal{S}$, as an upper bound for 3 separate sequences of Fourier coefficients, where $\mathcal{S}$ is the set of Schwartz functions. A sequence is chosen for each spatial coordinate, so tracks can be generated that can be parametrized $(f(t),g(t),h(t))$ for $t \in [0, 2 \pi]$, with periodic components with a period of $2 \pi$ and $f(0) = g(0) = h(0) = 0$. For each mode of oscillation, $n$, we randomly choose a $c_{n_x}$, $c_{n_y}$, and $c_{n_z}$ in $[0,s(n)]$. To maintain rotational invariance in our Fourier space, we reject all three and choose again if $c^{2}_{n_x}+c^{2}_{n_y}+c^{2}_{n_z} > s^{2}(n)$, so we can think of the three coefficients as being a single point chosen with a uniform distribution inside of some sphere of radius $s(n)$. Thus, we can think of a single track as generated by a sequence of points, where the nth point is chosen to lie inside a sphere of radius $s(n)$. If $f(t)$, $g(t)$, and $h(t)$ are differentiable with respect to time to all orders the track is called 'smooth'. 

To show that any smooth track can be generated by choosing the correct Schwartz function, suppose that $g(t) \in \mathbb{R}$ is smooth. We want to show that there exists a function $s \in \mathcal{S}$ for which the $C_n \le s(n)$ for every $n$, where the sequence $C_n$ is the Fourier representation of $g$. This can also be stated as: $$C_n = \frac{1}{2\pi}\int^{2 \pi}_{0} g(x) e^{inx}dx$$

We can then integrate by parts to show that 
$$C_n = -\frac{1}{2\pi}\frac{1}{n i}\int^{2 \pi}_{0} g'(x) e^{inx}dx$$

and if we now integrate by parts $k$ times, 
$$C_n = (-1)^{k}\frac{1}{2\pi}(\frac{1}{n i})^{k}\int^{2 \pi}_{0} g^{k}(x) e^{inx}dx$$

Therefore, it must be the case that 
$$\abs{\int^{2 \pi}_{0} g^{k}(x) e^{inx}dx} \le 2\pi \: \text{max}(g^{k}(x))$$

\noindent for $x \in [0, 2 \pi]$, and thus it can be immediately shown that 

$$\abs{C_n} = \abs{\frac{1}{2\pi}(\frac{1}{n i})^{k}\int^{2 \pi}_{0} g^{k}(x) e^{inx}dx} \le (\frac{1}{n})^{k} \: \text{max}(g^{k}(x)) $$

 Therefore $C_n$ is bounded from above by a Schwartz function, and every smooth track can be generated by careful choice of Schwartz function.

It is also important that this method \textit{only} generates smooth tracks regardless of choice of Schwartz function. This is easy to show, as the Fourier transform is an isomorphism onto the Schwartz space. This means that choosing a sequence of Fourier coefficients to lie on the Schwartz function: $C_n = s(n)$, makes the resulting function Schwartz. If instead: $C_n \leq s(n)$, then the resulting function is even more rapidly decaying, and is thus still Schwartz, and thus smooth.

It is illustrative to describe helical tracks in a Fourier basis and show that they are a subset of tracks that can be generated by the Schwartz function method described. In tracking, helices are typically fully described by the five parameters $(\phi_0, p_\textrm{T}, d_0, d_z, \tan(\lambda))$ where $\phi_0$ is the initial angle of the particle's trajectory with respect to the z axis, $p_\textrm{T}$ is the transverse momentum, $d_0$ is the distance of closest approach to the origin in the $x-y$ plane, $d_z$ is the distance of closest approach to the origin along the $z$ direction, and $\tan(\lambda)$ is the slope of the helix in the $r-z$ plane. The position of a particle for any time $t$ can then be described 

\[ \begin{cases} 
      x =  d_0 \cos{(\phi_0)} + \alpha \; p_\textrm{T} \{\cos{(\phi_0)} - \cos{(\phi_0 - \omega t)} \}\\
      y =  d_0 \sin{(\phi_0)} + \alpha \; p_\textrm{T} \{\sin{(\phi_0)} - \sin{(\phi_0 - \omega t)} \}\\
      z =  d_z - \alpha \; p_\textrm{T} \tan(\lambda) \; \omega t\\ 
   \end{cases}
\]
Where $\alpha=\frac{1}{qB}$ is the magnetic field constant.

This form can be expanded in a Fourier basis by first recognizing the form of the x and y components to already be trivially in a Fourier basis and expanding the $z$ component, 

$$ z =  d_z + \alpha \; p_\textrm{T} \tan(\lambda) \;  \omega t -  \sum_{n=1}^{\infty}\frac{2 \;\alpha \; p_\textrm{T} \tan(\lambda)}{n} \; \sin{(n \omega t)} $$

For any value of $t$ in a compact interval and any values of track parameters, there exists a Schwartz function that generates all the correct coefficients for this helical form. With the assumption that tracks start at the origin, the Schwartz functions chosen in this work will generate this form of helical track if $\alpha \; p_\textrm{T} \tan(\lambda))$ is not too large.

\subsection{Signal track generation}
\label{sig-gen}

Our signal track generation consists of producing a spatial curve via Fourier series, making a very large number of points on the curve which are evenly spaced by arc length, and recording the point closest to each layer that the curve passes through and labeling it a "hit". 

When producing the Fourier series, the fundamental wavelength is arbitrarily chosen to be equal in length to the detector radius. So long as sufficiently many frequencies are included in track generation, this will not limit the tracks that can be generated. However, this does put constraints on the tracks, such as those in Section \ref{subsec:base}, that are generated with a finite number of frequencies. When making this arbitrary choice it is only important that the fundamental wavelength is a similar length to the radius of the detector, so that the finite Frequency Tracks do not leave the detector too soon or too late. 

\end{document}